\documentclass[12pt]{iopart}
\usepackage{epsf}
\usepackage{epsfig}

\begin{document}

\title[Coupled maps]{Coupled logistic maps
and non-linear differential equations}
\vskip 10pt
\author{
Eytan Katzav$^{1}$ and Leticia F. Cugliandolo$^{1,2}$}
\address{
$^1$Laboratoire de Physique Th{\'e}orique de l'{\'E}cole Normale
Sup{\'e}rieure,
\\
24 rue Lhomond, 75231 Paris Cedex 05, France
\\
$^2$Laboratoire de Physique Th{\'e}orique  et Hautes {\'E}nergies, Jussieu,
\\
5\`eme {\'e}tage,  Tour 25, 4 Place Jussieu, 75252 Paris Cedex 05, France
}

\begin{abstract}
We study the continuum space-time limit of a periodic one dimensional
array of deterministic logistic maps coupled diffusively.  First, we
analyse this system in connection with a {\it stochastic} one dimensional
Kardar-Parisi-Zhang (KPZ) equation for confined surface
fluctuations. We compare the large-scale and long-time behaviour of
space-time correlations in both systems.  The dynamic
structure factor of the coupled map lattice (CML) of logistic units in
its deep chaotic regime and the usual $d=1$ KPZ equation have a
similar temporal stretched exponential relaxation. Conversely, the
spatial scaling and, in particular, the size dependence are very
different due to the intrinsic confinement of the fluctuations in the
CML.  We discuss the range of values of the non-linear parameter in
the logistic map elements and the elastic coefficient coupling
neighbours on the ring for which the connection with the KPZ-like
equation holds. In the same spirit, we derive a continuum partial
differential equation governing the evolution of the Lyapunov vector
and we confirm that its space-time behaviour becomes the one of
KPZ. Finally, we briefly discuss the interpretation of the continuum
limit of the CML as a Fisher-Kolmogorov-Petrovsky-Piscounov (FKPP)
non-linear diffusion equation with an additional KPZ non-linearity and
the possibility of developing travelling wave configurations.
\end{abstract}

\section{Introduction}
\label{sec:introduction}

Extended dynamical systems can be modelled with infinite dimensional
non-linear partial differential equations but also with coupled map
lattices (CMLs). The relation between the two and, in particular, their
behaviour in the spatio-temporal chaotic regime is a subject of great
interest~\cite{dyn-syst2}.

Whether local chaos can be modelled by stochastic noise is a very
deep question that has been addressed from different angles and
with a variety of analytical and numerical approaches.
Back in the 40's Von Neumann and Ulam proposed to use the logistic
map in its chaotic domain as a pseudo random number
generator~\cite{vonNeumann2}.  Except for a set of initial values
of measure zero, the iterates of the logistic map are distributed
according to $p(x) = [\pi^2 \, x(1-x)]^{-1/2}$ and the sequence
$T^{-1}(x)$ with $T(x) = \sin^2(\pi/2 x)$ is equidistributed
within the unit interval $[0,1]$. However, it was later realized
that the time series thus generated has a number of defects, such
as time-correlations and a trivial first return-map; the
logistic map was subsequently neglected as a pseudo-random number
generator~\cite{discussion,Gonzalez}.

The question above was later revisited in the context of extended
dynamical systems modelled with either discrete models or continuous
partial differential equations. CMLs with
short-range spatial interactions and chaotic dynamics of the
independent units develop collective behaviour and long-range order in
finite regions of the phase diagram characterised by the strength of
the non-linearity and the coupling between the dynamic
units~\cite{Kaneko}.  Some statistical mechanics notions were shown to
be useful to describe their spatiotemporal
behaviour~\cite{dyn-syst0}-\cite{Marcq-etal}.

On the continuum side,
in~\cite{Yakhot}-\cite{Kapral}
the authors compared the deterministic, driven by inherent
instabilities, Kuramoto-Sivashinsky (KS) partial differential equation
equation~\cite{Kuramoto,Sivashinsky} to the stochastic
Kardar-Parisi-Zhang (KPZ) partial differential
equation~\cite{KPZ}-\cite{Barabasi-Stanley} for
surface growth. A careful study of the scaling properties of the
solution of both equations showed that they are identical in $d=1$ but
differ in $d\geq 2$~\cite{Procaccia-etal2,Procaccia-etal3},
a result confirmed numerically in~\cite{Krug}.

In this paper we investigate the spatio-temporal behaviour of a one
dimensional ring of coupled logistic maps by taking a formal continuum
space-time limit. First, we identify two non-linear terms as leading
to effective noise and confinement and we argue that the CML
has a long wave-length, long time behaviour that is very similar to
the one of a modified KPZ equation describing surface fluctuations in
a confining potential. Second, we revisit the
equation for the evolution of the Lyapunov vector using similar
arguments~\cite{Politi}.
Finally, we reckon that the resulting partial differential equation is
an extension of the Fisher-Kolmogorov-Petrovsky-Piscounov
(FKPP)~\cite{Fisher} non-linear diffusion equation with an additional
KPZ-like non-linearity and we exhibit some travelling wave
solutions.

The article is organized as follows. In Section~\ref{sec:coupledmaps}
we recall the definition of the logistic map and the CML model we
study.  In Section~\ref{sec:KS-KPZ} we review the partial differential
equations we are concerned with. We first describe the KS and KPZ
partial differential equations and we recall the connection between
them in $d=1$. We then present the FKPP equation and some properties
of its travelling wave solutions.  In these first sections we also set
the notation and we explain our goal.  In Section~\ref{sec:continuum}
we define the continuum limit and we discuss each term in the
resulting partial differential equation.  In Section~\ref{sec:results}
we present the results of the numerical integration of the CML. We
test the `noise' term, we confront the behaviour of several
observables to the corresponding ones in KPZ one dimensional growth,
and we discuss the behaviour of the Lyapunov vector.
Section~\ref{sec:travelling} is devoted to a brief discussion of the
travelling wave solutions.  Finally, in Section~\ref{sec:conclusions}
we present our conclusions and we discuss several proposals for future
research.

\section{Coupled logistic maps}
\label{sec:coupledmaps}

\subsection{The logistic map}

The logistic map is a non-linear evolution equation
acting on a continuous variable $x$ taking values in the
unit interval $[0,1]$. The evolution is defined
by iterations over discrete time, $n=0,1,\dots $:
\begin{equation}
x_n = f(x_{n-1}) \equiv r \, x_{n-1} (1-x_{n-1})
\; ,
\label{eq:logistic-map}
\end{equation}
with the parameter $r$ taking values $0 < r \leq 4$. The time
series has very different behaviour depending on the value of $r$.
For $0\leq r<1$ the iteration approaches the fixed point $x^*=0$.
For $1\leq r < 3$ the asymptotic solution takes the finite value
$x^*(r)=1-1/r$ for almost any initial condition. Beyond $r=3$ the
asymptotic solution bifurcates, $x_n$ oscillates between two
values $x_1^*$ and $x_2^*$, and the solution has period $2$.
Increasing the value of $r$ other bifurcations appear at sharp
values, {\it i.e.} further period doubling takes place. Very
complex dynamic behaviour arises from the relatively simple
non-linear map (\ref{eq:logistic-map}) in the range $r \in
[3.57,1]$: the map has bands of chaotic behaviour, {\it i.e.}
different initial conditions exponentially diverge, intertwined
with windows of periodic behaviour.  Surprisingly enough, some
exact solutions are known for special values of
$r$~\cite{Wolfram}. For example
$x_n = \sin^2(\theta \pi 2^n)$
with $k$ integer and $\theta$ determined by the initial condition
through $x_0=\sin^2(\theta\pi)$ is a solution for $r=4$
~\cite{vonNeumann}.

\subsection{Coupled map lattices}

Coupled map lattices (CMLs)~\cite{Kaneko} are discrete arrays of
scalar variables taking continuous values, typically in the unit
interval, that evolve over discrete time according to a dynamic rule.
They are generalisations of cellular automata for which the variables
take only discrete values. Kaneko introduced them as phenomenological
models to describe media with high energy pumping but they may also
arise as discrete versions of partial differential equations.

A typical realization of the evolution of a single independent
unit is given by the logistic map (\ref{eq:logistic-map}). One
usually uses models in one or two space dimensions with periodic
boundary conditions (a ring or a torus). A Laplacian coupling
among nearest-neighbours on the lattice is usually chosen as it is
motivated by the intent to model fluid mechanics in a simpler
manner.  In $d=1$, and labelling with $i=1,\dots, N$ the sites on
the ring, the coupled map model takes the form
\begin{equation}
x^i_n = f(x_{n-1}^i) +
\frac{\nu}{2} \,
[ \, f(x_{n-1}^{i-1}) - 2 f(x_{n-1}^{i})  + f(x_{n-1}^{i+1}) \, ]
\label{eq:coupled-maps}
\end{equation}
with $x_n^{i+N}=x_n^i$ for all $n$ with $N$
the number of elements on the ring.
The initial condition is usually chosen to be random and thus taken from
the uniform distribution on the interval $[0,1]$ independently on each site.
$\nu$ is the coupling strength between the nodes and plays the role
of a viscosity.

Numerical simulations showed that CMLs exhibit a large variety of
space-time patterns: kink-antikink configurations, space-time periodic
structures, wavelike patterns or spatially periodic structures with
steady, periodic or chaotic dynamics, space-time intermittent and
spatio-temporal chaos have been found for different values of the
parameters ($r$ and $\nu$). A detailed description of the phase
diagram is given in~\cite{Kaneko}.  In a nutshell,
the dynamics is characterised by a competition between the diffusion
term, that tends to produce an homogeneous behaviour in space, and the
chaotic motion of each unit, that favours spatial
inhomogeneous behaviour due to the high sensitivity to the initial
conditions.

\section{Partial differential equations}
\label{sec:KS-KPZ}

The $d=1$ Kuramoto-Sivashinsky (KS) equation was introduced by Kuramoto to
study local phase turbulence in cyclic chemical
reactions~\cite{Kuramoto}.  A $d=2$ extension of it was later used by
Sivashinsky to study the propagation of flame fronts in mild
combustion~\cite{Sivashinsky}. In $d=1$ it reads
\begin{equation}
\frac{\partial h}{\partial t} = -\kappa h - \frac{\partial^2
h}{\partial x^2} - \frac{\partial^4 h}{\partial x^4} - h
\frac{\partial h}{\partial x} \label{eq:KS}
\end{equation}
with $h=h(x,t)$ a real function and $\kappa$ a real parameter.
Similar pattern formation to the one developed in CMLs has been
observed in the numerical solutions of the KS
partial differential equation~\cite{Frisch}.

The Kardar-Parisi-Zhang (KPZ) equation was proposed as a non-linear model
that describes surface growth~\cite{KPZ}. If $h(x,t)$ is the height of a
surface on a substrate point $x$ at time $t$, the equation reads
\begin{equation}
\frac{\partial h}{\partial t} = \nu \frac{\partial^2 h}{\partial
x^2} +\frac{\lambda}{2} \left( \frac{\partial h}{\partial
x}\right)^2 + \eta(x,t)
\label{eq:KPZ}
\end{equation}
with $\eta(x,t)$ a Gaussian white noise with zero mean and
correlation given by
\begin{equation}
\langle\, \eta(x,t) \eta(x',t')\, \rangle = D \, \delta(x-x') \, \delta(t-t')
\; .
\label{eq:white-noise}
\end{equation}
The variable $dh/dx$ satisfies a noisy Burgers
equation~\cite{Burgers}. (For a discussion of the properties of
these equations see ref.~\cite{Barabasi-Stanley}.)

Using a dynamic renormalisation group
calculation, Yakhot suggested that the elimination of large wave-vector
modes generates a random `stirring force' with zero mean and average
$\langle \, f_i(\vec k,\omega) f_j(\vec k', \omega') \, \rangle = k_i
k_j \, \delta(\vec k+\vec k') \, \delta(\omega+\omega')$.
The average
$\langle \dots \rangle$ represents a time-average in the stationary
state.  Simplifying further the propagator in the $\vec k\to 0$ limit,
he argued that in the large scale and long time limit, $\vec k\to
0$ and $\omega\to 0$, the KS equation becomes the random-force-driven
Burgers equation or, under a change of variables, the
KPZ equation (with positive viscosity in $d=1$
and negative viscosity in $d \geq 2$)~\cite{Yakhot}.

Next came a number of papers in which  L'vov, Procaccia {\it et
al}~\cite{Procaccia-etal1}-\cite{Procaccia-etal3}
studied this problem in more detail.
They showed that for both equations the field-field correlation in
Fourier space,
$n(\vec k,\omega)$, defined from
\begin{equation}
\langle \, h(\vec k, \omega) h^*(\vec k',\omega') \, \rangle
=
n (\vec k, \omega) \, \delta(\vec k-\vec k') \, \delta(\omega-\omega')
\; ,
\end{equation}
has a scale invariant form
\begin{equation}
n(\vec k,\omega) = \frac{n(\vec k)}{\nu k^z} \;
f\left( \frac{\omega}{\nu k^z} \right)
\; ,
\qquad
n(\vec k) = \int_\omega n(\vec k,\omega) = \frac{n}{k^{2\alpha+d}}
\; ,
\label{eq:scale-inv}
\end{equation}
with the `roughness exponent' $\alpha$ and the `dynamic exponent' $z$
being equal to $1/2$ and $3/2$ in $d=1$,
respectively~\cite{Procaccia-etal1}.  Under the assumption that a
scale invariant solution of the form (\ref{eq:scale-inv}) also exists
in $d>1$, L'vov and Procaccia~\cite{Procaccia-etal2}
argued that the scaling solutions for
the two models bifurcate in $d=2$; that is to say, the exponents $z$
and $\alpha$ are not the same in {\it all} dimensions and hence the
two equations do not belong to the same universality class.
Further evidence for the breakdown of the relation between KS and
the strong coupling phase of KPZ in $d\geq 2$ appeared in
\cite{Procaccia-etal3}.

It has been difficult to verify the coincidence of the scaling
solution to the KS and KPZ equations in $d=1$ by solving the KS
equation numerically and comparing to the prediction $z=3/2$.
The reason for this difficulty is the existence of a very long
crossover regime~\cite{Zaleski}.  A careful numerical
study appeared in \cite{Krug}
where large scale simulations were confronted
to the predictions for the crossover behaviour obtained from the
analysis of the KPZ equation.

The Fisher-Kolmogorov-Petrovsky-Piscounov (FKPP) non-linear diffusion
equation determines the evolution of the concentration of some chemical species
or individuals, $0\leq h \leq 1$, on a one dimensional space and reads
\begin{equation}
\frac{\partial h}{\partial t}=
\nu
\frac{\partial^2 h}{\partial x^2} + f(h)
\; ,
\label{eq:FKPP}
\end{equation}
with $f(h)$ a non-linear term, typically
of the form $f(h) = k_1 h -k _2 h^2$ with $k_1>0$ and $k_2>0$.
The FKPP equation has travelling wave solutions~\cite{vanSaarloos}
\begin{equation}
h(x,t) = F(x-vt)
\label{eq:travel}
\end{equation}
that represent the invasion of the stable
phase $h(-\infty,0)=k_1/k_2$  in the
unstable phase $h(\infty,0)=0$
and travel with velocity
$v\geq 2 \sqrt{k_1}$.  Which velocity is selected depends
on the initial condition. In many cases, if
the initial front is ``sufficiently steep'', {\it i.e.} $F(y)$ decays
faster than $e^{-\gamma_{min} y}$ for $y \to \infty$~\footnote{a step
function realizes this requirement} the front advances asymptotically
with the minimal velocity $v_{min}$.  These are called ``pulled
fronts'': the leading edge of the front pulls the interface through
growing linear perturbations about the unstable $h=0$ value. In these
cases, the velocity approaches $v_{min}$ as a power law
\begin{equation}
v(t) = v_{min} - \frac{3}{2 \gamma_{min} t} + {\cal O}(t^{-3/2})
\; .
\end{equation}
In ``pushed fronts'' instead it is the non-linear growth in the region behind
the leading edge that pushes the interface and the asymptotic velocity is
larger than $v_{min}$.

\section{The continuum limit of the CMLs}
\label{sec:continuum}

In this Section we define the continuum limit and we briefly discuss
each term in the resulting partial differential equation  by
making an explicit comparison to the ones in the KPZ equation
(\ref{eq:KPZ}) and the FKPP equation (\ref{eq:FKPP}).


\subsection{The continuum limit}

The main idea is to take the continuum limit of the CML
using the following discretisation of time and space
derivatives:
\begin{eqnarray}
\frac{\partial h}{\partial t}
\;\;\;
&\leftrightarrow&
\;\;\;
\frac{h_{n+1}^i - h_n^i}{\delta t}
\; ,
\label{eq:continuum1}
\\
\frac{\partial h}{\partial x}
\;\;\;
&\leftrightarrow&
\;\;\;
\frac{h^{i+1}_n-h_n^i}{\delta x}
\; ,
\label{eq:continuum2}
\\
\frac{\partial^2 h}{\partial x^2}
\;\;\;
&\leftrightarrow&
\;\;\;
\frac{h_n^{i+1}-2 h_n^i+h_n^{i-1}}{(\delta x)^2}
\; ,
\label{eq:continuum3}
\\
\frac{\partial^4 h}{\partial x^4}
\;\;\;
&\leftrightarrow&
\;\;\;
\frac{h_n^{i+2}-4 h_n^{i+1}+6 h_n^i-
4 h_n^{i-1}+h_n^{i-2}}{(\delta x)^4}
\; ,
\label{eq:continuum4}
\end{eqnarray}
with $\delta t$ the time-step and $\delta x$ the lattice spacing
equal one in our system of units.
The CML of logistic elements then becomes
\begin{eqnarray}
\frac{\partial h}{\partial t} &=& \frac{\nu r}{2} (1-2 h) \; \frac{\partial^2
h}{\partial x^2} - \nu r\left( \frac{\partial h}{\partial x}\right)^2 +
(r-1) h - r h^2
\; ,
\label{eq:CML}
\end{eqnarray}
where we called $x$ the coordinate ($i \delta x \to x$),
$t$ the time $(n \delta t \to t$),
and $h$ the field [$x_n^i\to h(x,t)=h$].

\subsection{Relation with KPZ}

Comparing to eq.~(\ref{eq:KPZ}) one notices that:

\begin{enumerate}

\item[(i)] By definition the field $h$ is bounded and takes values in the unit
interval. Thus, the resulting equation should have an effective
confining potential that limits the field to a finite range.

\item[(ii)] The elastic term is here multiplied by a
field-dependent viscosity
\begin{equation}
\nu(h) \equiv \frac{\nu r}{2} (1-2h)
\; .
\end{equation}

\item[(iii)] The second, non-linear term is of the form of the one
in the KPZ equation with a negative coupling
\begin{equation}
\lambda \equiv -\nu r
\; .
\end{equation}

\item[(iv)] The last two terms read
\begin{equation}
\eta(x,t) \equiv (r-1) h(x,t) - r h^2(x,t)
\; .
\label{eq:eta}
\end{equation}

We notice that these terms are not present in the KPZ equation.
In order to  compare to the latter we shall argue that they have
a double identity: on the one hand $\eta$ behaves roughly as a
short-range correlated noise in space and time; on the other hand
it can be interpreted as a force derived from a confining
potential
\begin{equation}
\eta = -\frac{\partial V(h)}{\partial h}
\; ,
\qquad \qquad
V(h) = - \frac{(r-1)}{2} h^2 + \frac{r}{3} h^3
\; .
\label{eq:Vh}
\end{equation}

\end{enumerate}

In the next subsections we briefly discuss the elastic and
non-linear terms.  The analysis of $\eta$ is more
delicate and we postpone it to the next section where we present
the results of the numerical integration of the CML.

\subsubsection{The elastic term.}

The fact that the elastic term is multiplied by a field-dependent
viscosity needs a careful inspection.

First,  $\nu(h)$ is negative for $h<1/2$ which implies an
instability in the hydrodynamic limit. The same feature appears in
the KS equation [eq.~(\ref{eq:KS})]. It was shown that this
instability taps the system and so creates an effective `noise'
leading to the mapping of the KS equation onto a similar equation
with a noise term and a renormalised positive viscosity
coefficient~\cite{Yakhot,Procaccia-etal1}
(see the discussion in Sect.~\ref{sec:KS-KPZ}).  With
respect to this `noise creation' feature, it is also very
important to have a confining potential which restraints the
instabilities caused by the negative values of $\nu(h)$.
Together they create the effective noise. Therefore, the fact
that we may get negative values for the viscosity actually
supports our line of argumentation, {\it i.e.} that of relating a
deterministic model to a stochastic one.

Second, if $h$ remains bounded between, say, 0 and 1 the `bare'
viscosity takes values on the finite interval
$[-\frac{\nu r}{2},\frac{\nu r}{2}]$.
However, as is well known, in the KPZ system the viscosity
coefficient is renormalised by the nonlinear term (put in simple
words, the nonlinear term has a smoothing effect) so that the
large-scale viscosity that the system experiences is not only
determined by the bare value. Thus, we also expect the
field-dependent bare viscosity to be renormalised at large scales
and thus its precise value not to be very important.

\subsubsection{The non-linear term.}

It is well-known that the sign of the coupling constant is not
important in the KPZ equation. The reason is that an inversion of
the sign of $\lambda$ just corresponds to
describing the evolution of the mirror image of the original
surface ({\it i.e.} a description of $-h(x,t)$, see
{\it e.g.}~\cite{Barabasi-Stanley}). Actually, when mapping the noisy
Burgers equation onto the KPZ equation the resulting coupling
constant is negative, namely $\lambda=-1$.

\subsection{Relation with FKPP}

Comparing to eq.~(\ref{eq:FKPP}) one realizes that

\begin{enumerate}

\item[(i)] The field $h$ is bounded as in FKPP.

\item[(ii)] The viscosity is now field dependent
and may take negative values.

\item[(iii)] There is a KPZ-like non-linearity, not present in the
FKPP equation.

\item[(iv)] The last two terms are identical to the ones in the FKPP
equation with $k_1=r-1$ and $k_2=r$.

\end{enumerate}

The similarity with the FKPP equation suggests to search for
travelling wave solutions in the CML and study their properties
(velocity, etc.) as done for the FKPP equation~\cite{vanSaarloos}.
We shall come back to this issue in Sect.~\ref{sec:travelling}.

\section{Numerical tests in the context of surface growth}
\label{sec:results}

We integrated numerically the CML of logistic units
with up to $N=1024$ sites and periodic boundary conditions. All units
were updated in parallel. We used
floating point precision $10^{-16}$. Unless otherwise stated we
chose independent random initial conditions for each map taken
from the unit interval with flat probability. Since finite-size
effects seem to be negligible for systems larger than
approximately $32$ sites~\cite{Mousseau},
we do not discuss smaller systems, nor
do we make a systematic study of finite-size scaling. In
particular, we discuss the choice of parameters in
Sect.~\ref{choice}, we study the statistical properties of the
$\eta$ term in Sect.~\ref{noise}, we compare the spatial and
temporal behaviour of correlation functions in the CML and
KPZ growth in Sect.~\ref{sec:confront}, and we
analyse the Lyapunov vector in Sect.~\ref{sec:Lyapunov}.

\subsection{Choice of parameters}
\label{choice}

\begin{figure}[htb]
\includegraphics[width=5cm]{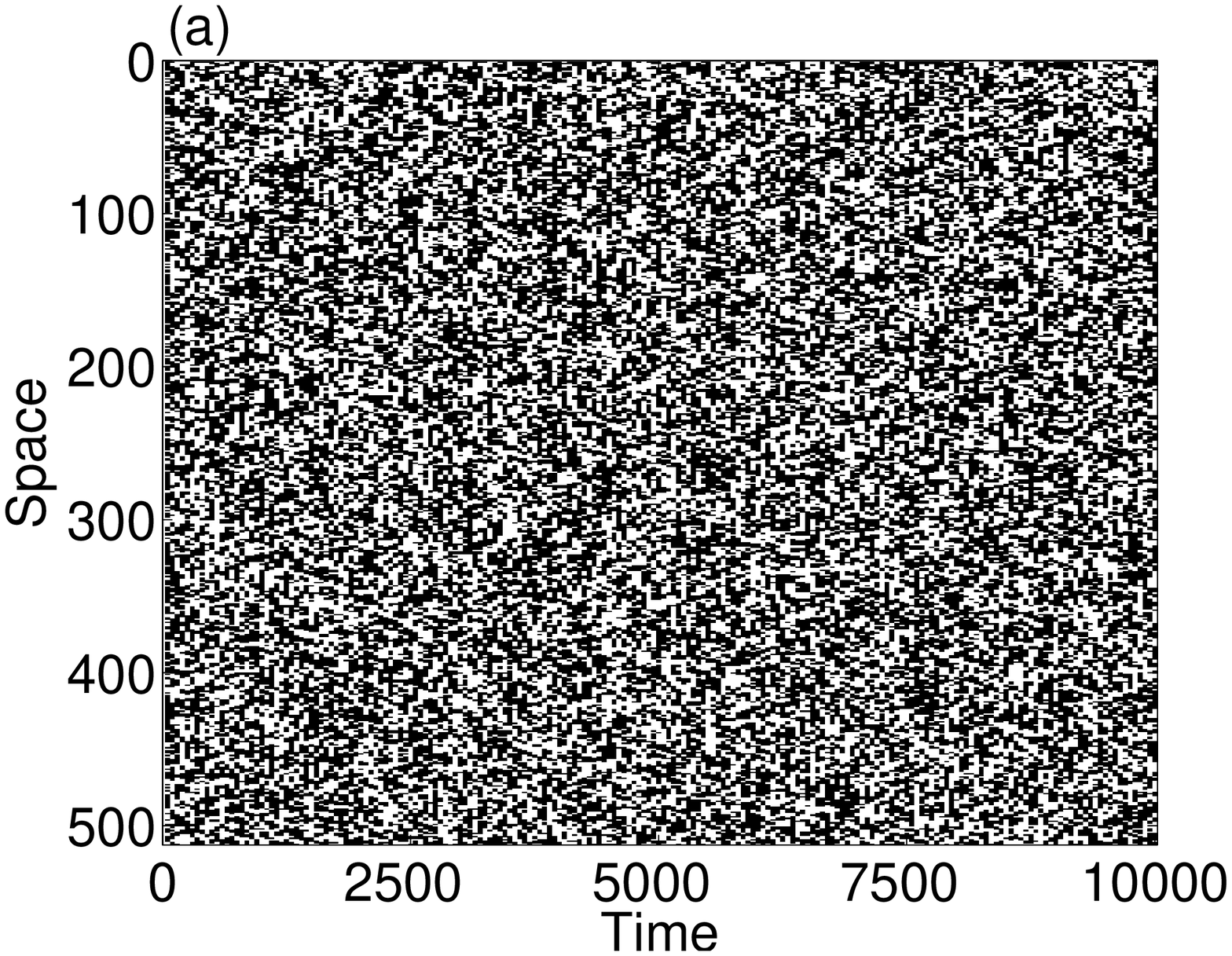}
\includegraphics[width=5cm]{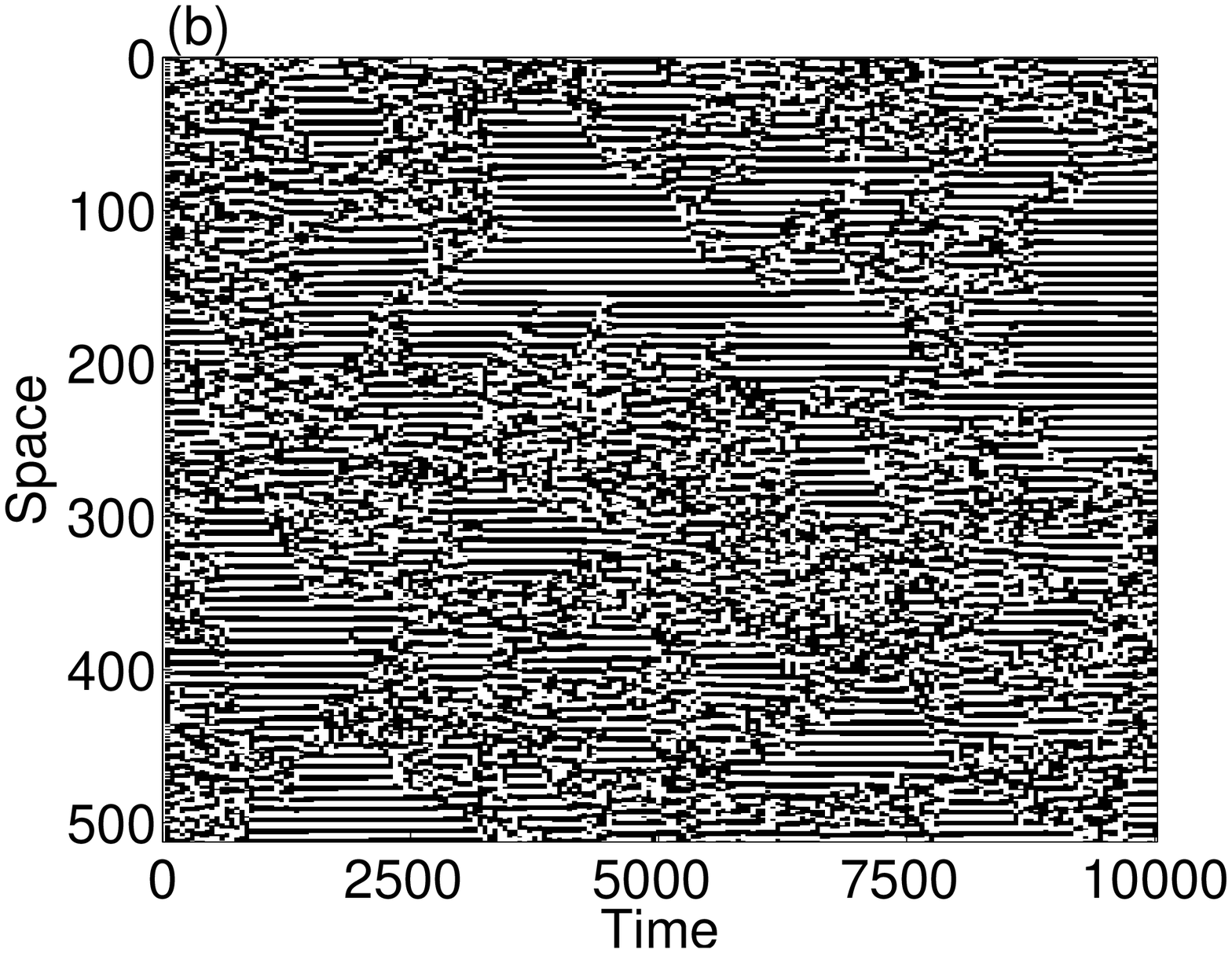}
\includegraphics[width=5cm]{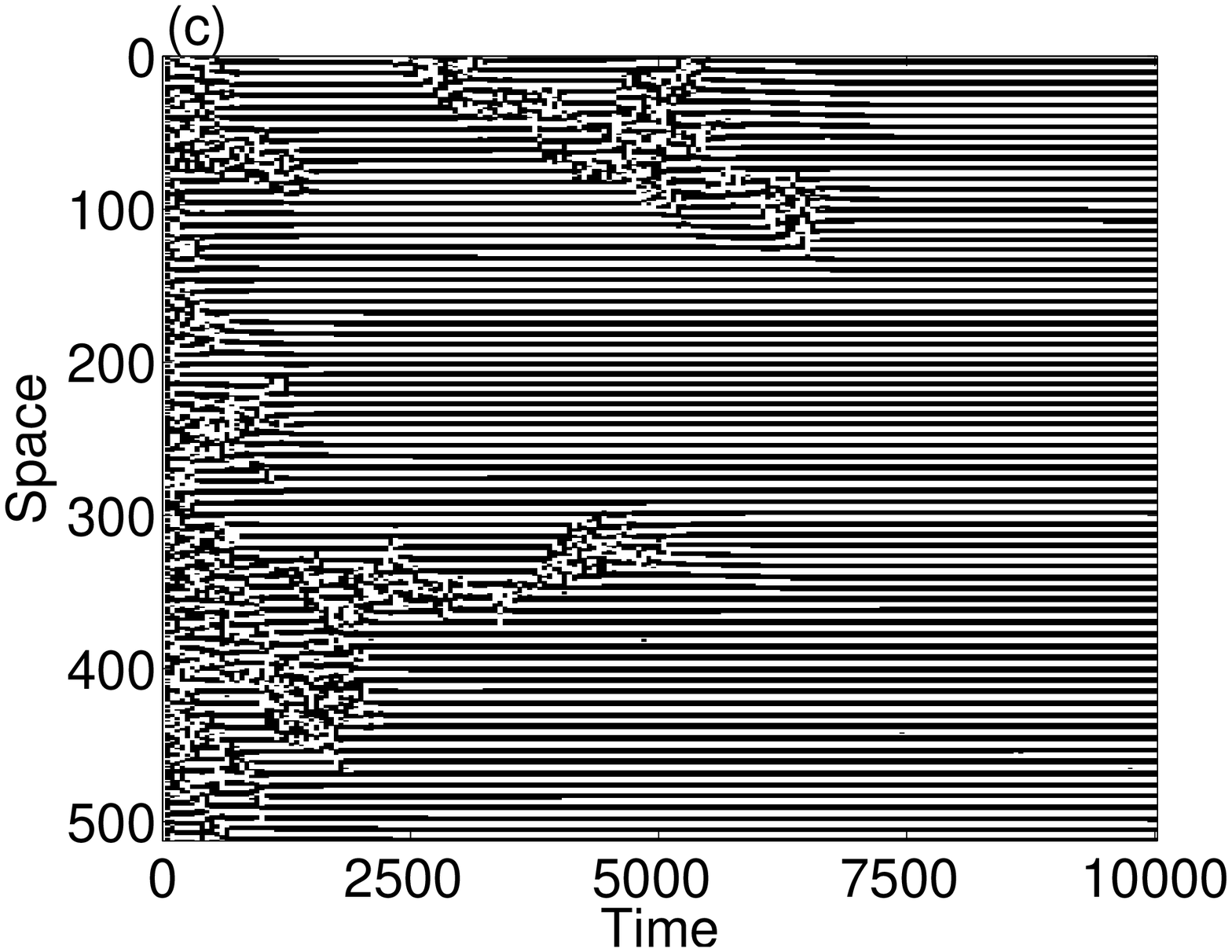}
\caption{Space-time plots of the coarse-grained variable
$s_n^i = \mbox{sign}(x_n^i-x^*)$ for (a)
$r=4$, (b) $r=3.82$, and (c) $r=3.8$.}
\label{figa}
\end{figure}

The panels in Fig.~\ref{figa} present the space-time plot of a
coarse-grained variable obtained by transforming the continuous
variable $x_n^i$ into a bi-valued Ising-like one $s_n^i =
\mbox{sign}(x_n^i-x^*)$ for several values of the non-linear
parameter $r$ and the coupling strength $\nu=0.4$. We draw time on
the horizontal axis and space on the vertical one. Every $50$ time
step is plotted; if $x_n^i$ is larger than $x^*=1-1/r$ (the
unstable fixed point of a single logistic map) the corresponding
pixel is painted black; otherwise it is left white. This analysis
allows us to identify different regions of phase
space~\cite{Kaneko} in which we can study the connection with
the KPZ equation. In this work we focus on $\nu=0.4$ and $r=4$,
{\it i.e.} deep in the chaotic regime. We briefly mention at the
end of this Section the behaviour found for other values of $\nu$
and $r$.

\subsection{Analysis of $\eta(x,t)$}
\label{noise}

\subsubsection{The noisy aspect.}

\begin{figure}[h]
\includegraphics[width=7cm]{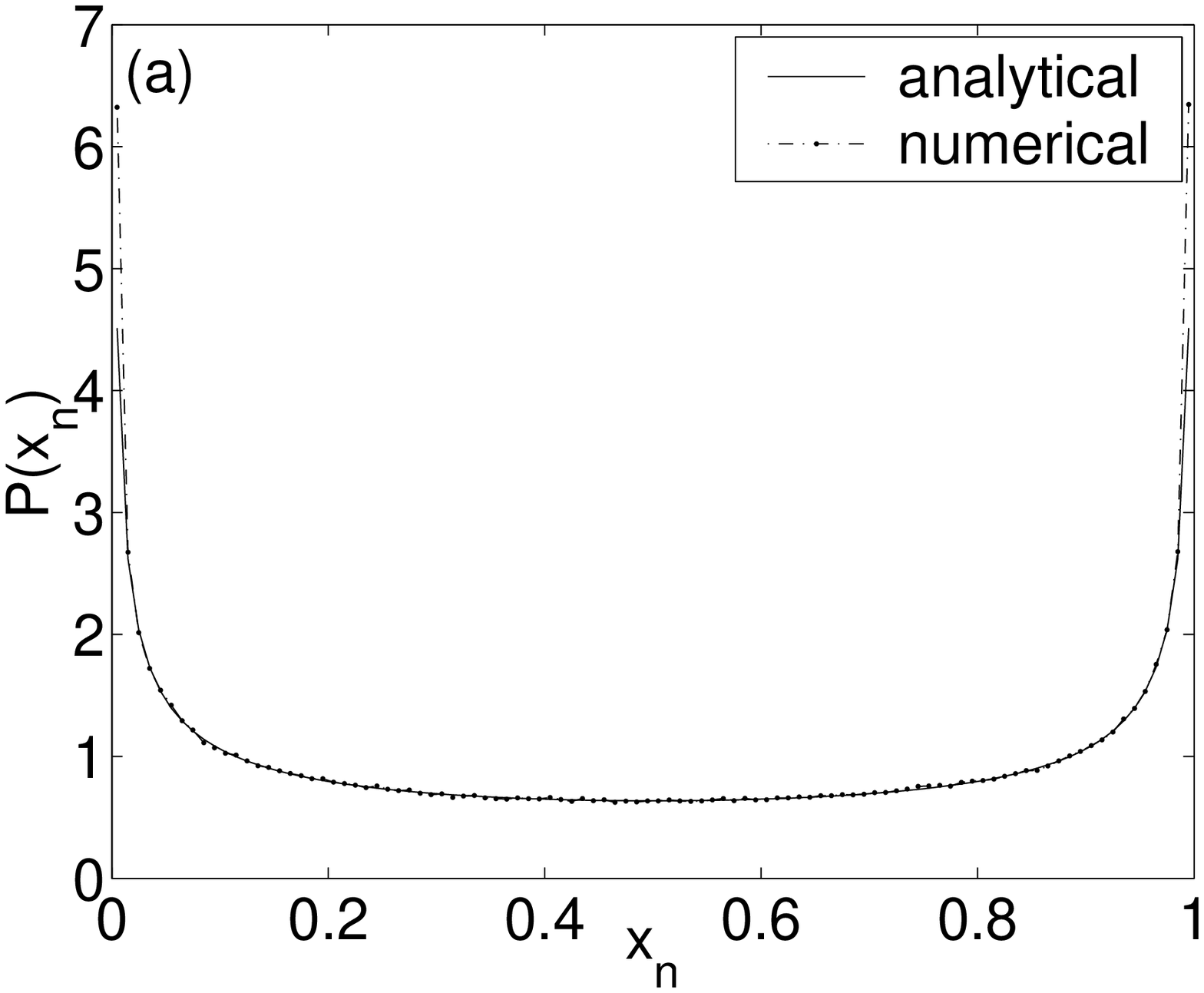}
\includegraphics[width=7.5cm]{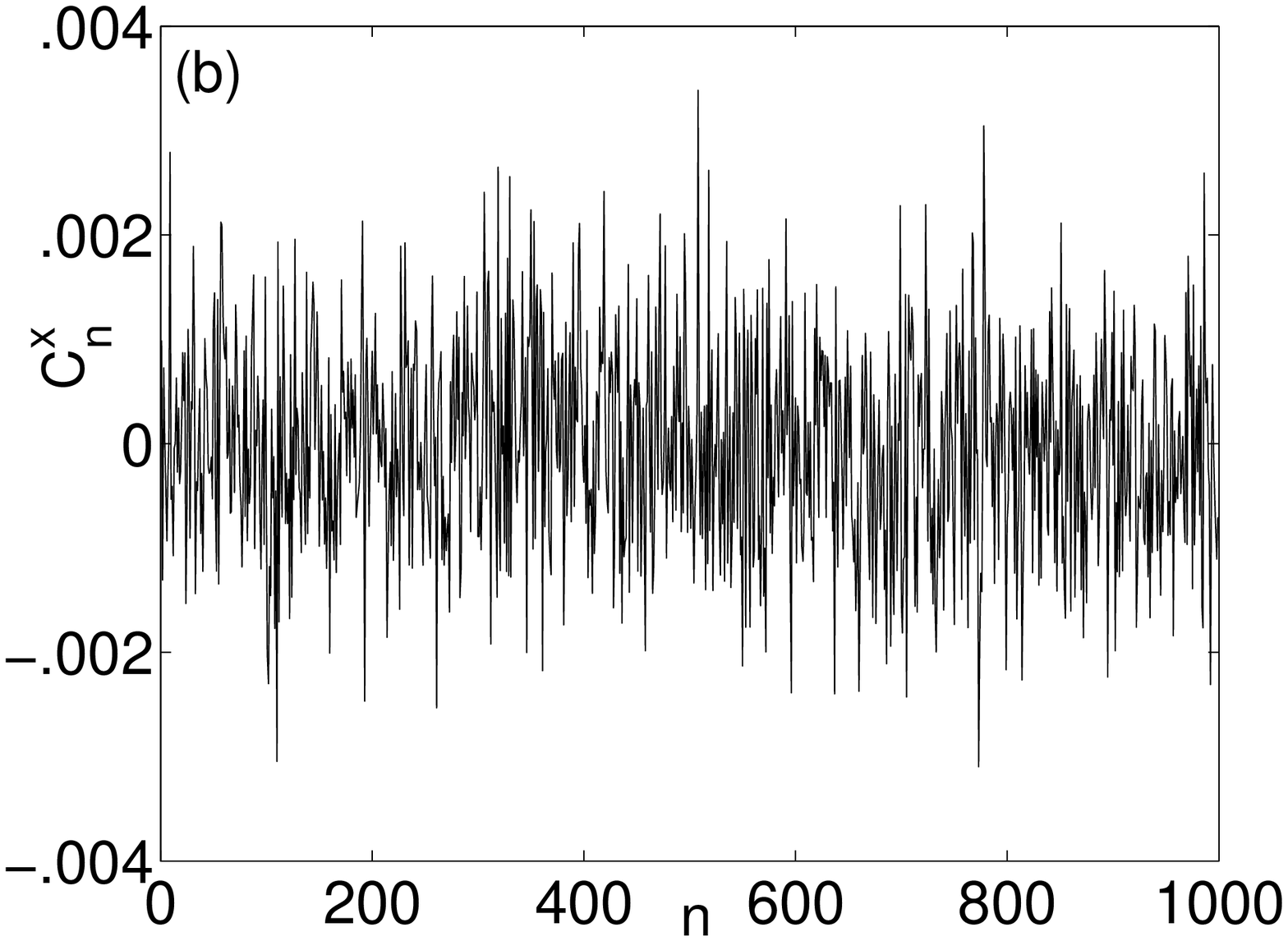}
\caption{(a) The PDF of a single map $x_n$, and (b) the evolution of
its autocorrelation function $C_n^x$ defined in
eq.~(\ref{eq:Cn-def}).}
\label{singleMAP}
\end{figure}

Before studying the properties of $\eta$ in the CML let us recall some
properties of this quantity in the single map. Choosing the parameter
$r$ to be $r=4$, {\it i.e.} well in the chaotic regime, we obtain the
histogram of $x_n$ over a time window of length $T=10^6$ shown in
Fig.~\ref{singleMAP}(a). The figure demonstrates that the numerical precision
chosen is good enough for our purposes since the data are rather
well described by the analytic prediction for the probability distribution
function (PDF)
\begin{equation}
p(x) = \frac{1}{\pi\sqrt{x(1-x)}}
\; .
\label{eq:px}
\end{equation}
The divergence of the peaks at $x=0$ and $x=1$ is however suppressed
due to the floating point precision
of the numerical data.
The histogram is symmetric around $x=1/2$ and,
consequently, the mean is given by $\langle \, x \, \rangle\equiv
T^{-1} \sum_{n-1}^T x_n = 1/2$.

In Fig.~\ref{singleMAP}(b) we show
the connected correlation over time,
\begin{equation}
C_n^x \equiv
\langle\, \Delta x_{k+n} \Delta x_k \, \rangle
=
\frac{1}{T-n} \sum_{k=0}^{T-n} \Delta x_{k+n} \Delta
x_k \label{eq:Cn-def} \; ,
\end{equation}
for a single run using a randomly chosen seed. $\Delta x_k
\equiv x_k-1/2$ and the average is taken over all pairs of data
obtained with a time delay $n$ on a single run of length $T=10^6$. No
correlations can be identified from this plot
though we know, however, that some
time-structure might be present~\cite{Wagner}
(see, {\it e.g.} the return map in Fig.~\ref{return}).

\begin{figure}[htb]
\includegraphics[width=7cm]{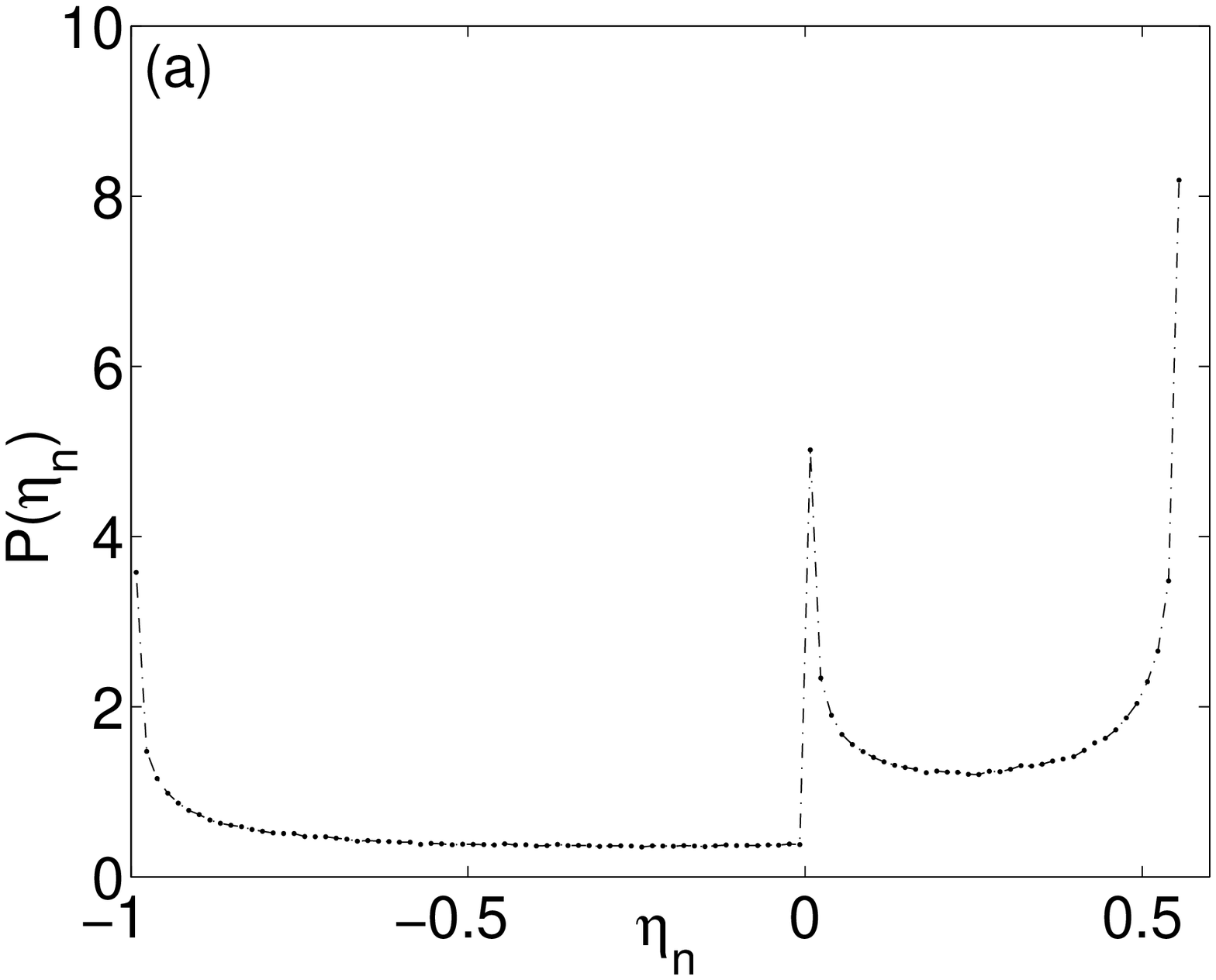}
\includegraphics[width=7.5cm]{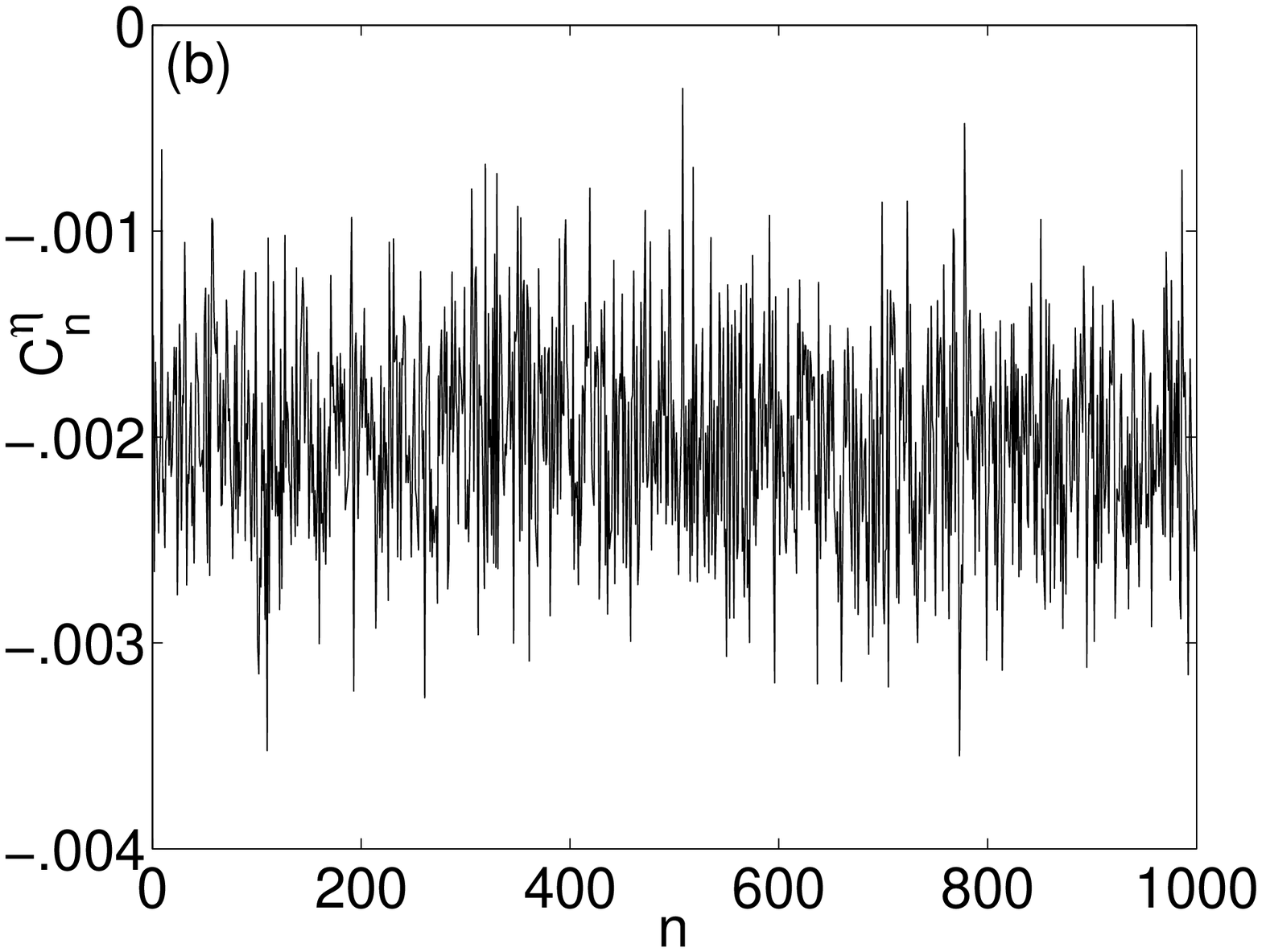}
\caption{(a) The PDF of the `noise'
term in a single map $\eta_n$. (b) The evolution of
its connected autocorrelation
function, $C_n^\eta$, over time.}
\label{fig2}
\end{figure}

We now turn to the study of
\begin{equation}
\eta_n \equiv (r-1) x_n - r x_n^2
\; .
\end{equation}
By definition, $\eta_n$ takes values in the interval
$[-1,(r-1)^2/(4r)]$ that for $r=4$ becomes $[-1,9/16]$. In
Fig.~\ref{fig2}(a) we show the histogram of $\eta_n$
corresponding to the same data shown in Fig.~\ref{singleMAP}.
Actually, using the PDF of
$x_n$ in eq.~(\ref{eq:px})
we derive the PDF of $\eta_n$ that is given by
\begin{eqnarray}
  p\left( \eta  \right) &=& \frac{1}
{{2\pi \sqrt {\frac{3} {4} - \sqrt {\frac{9} {{16}} - \eta } }
\sqrt {\frac{5} {4} + \sqrt {\frac{9} {{16}} - \eta } } \sqrt
{\frac{9}
{{16}} - \eta } }} \; \theta \left( \eta  \right) \nonumber \\
   && + \frac{1}
{{2\pi \sqrt {\frac{3} {4} + \sqrt {\frac{9} {{16}} - \eta } }
\sqrt {\frac{5} {4} - \sqrt {\frac{9} {{16}} - \eta } } \sqrt
{\frac{9} {{16}} - \eta } }}
\; .
\end{eqnarray}
where $\theta (\eta)$ is the Heaviside (step) function. The
numerical histogram is well-described by this analytic function
with the proviso that the divergences are replaced by the peaks in
the figure. Using this expression for $p(\eta)$ one finds that the
average of $\eta$ vanishes
\begin{equation}
\langle \, \eta \, \rangle = (r-1) \langle \, x \, \rangle - r
\langle \, x^2\,\rangle = 3 \times \frac12 - 4 \times \frac38 = 0
\; .
\end{equation}
A vanishing mean is recovered by computing the average numerically.
Similarly we get
\begin{eqnarray}
\mbox{Skewness} =
\frac {\langle \eta^3 \rangle} {\langle \eta^2 \rangle^{3/2}}
= - \frac {3} {4} \; ,
\qquad
\mbox{Kurtosis} =
\frac {\langle \eta^4 \rangle} {\langle \eta^2
\rangle ^2}=\frac{9}{4}
\; .
\end{eqnarray}

The time dependence of the connected autocorrelations,
$C^\eta_n \equiv
\langle \, \Delta \eta_{k+n} \Delta\eta_k \, \rangle
\equiv
{(T-n)}^{-1} \sum_{k=0} \Delta \eta_{k+n} \Delta\eta_k =$$
{(T-n)}^{-1} \sum_{k=0} \eta_{k+n} \eta_k$,
are shown in Fig.~\ref{fig2}(b) and again no obvious structure is observed.

\begin{figure}[htb]
\includegraphics[width=7cm]{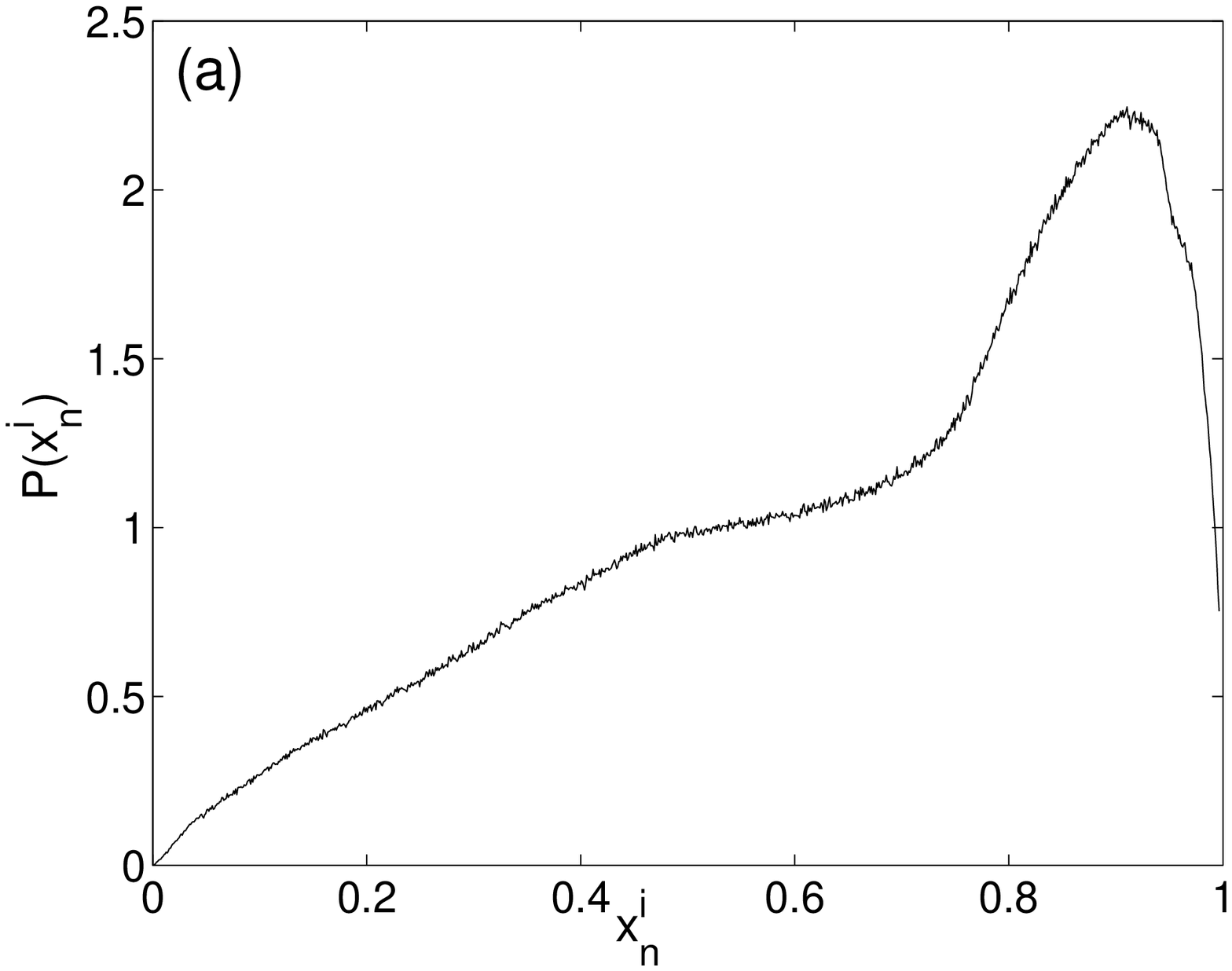}
\includegraphics[width=7cm]{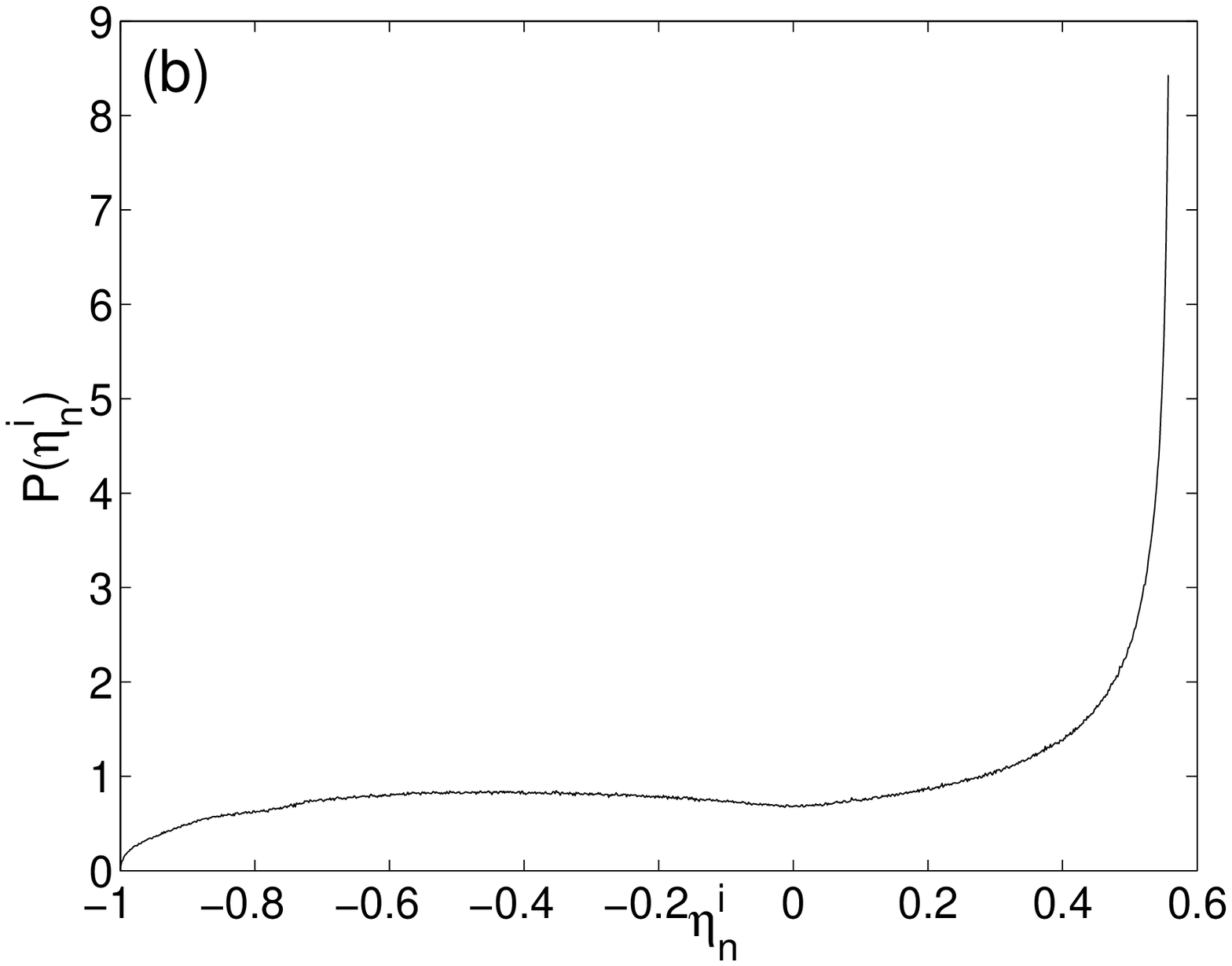}
\caption{The PDF of (a) a typical map $x^i_n$ and (b) a typical
noise term in the fully chaotic regime ($r=4$).}
\label{fig3}
\end{figure}

Next, we study the noise term in the CML.  In Fig.~\ref{fig3}(a) we
show a representative histogram of the individual units $x_n^i$. This PDF
is clearly smoother than the one of the independent map
shown in Fig.~\ref{singleMAP}(a).
For
the rather high value of the coupling strength used, $\nu=0.4$, all
units behave statistically in the same way. The average is
$\langle \, x \, \rangle \approx 0.67$.
In Fig.~\ref{fig3}(b) we show a representative
histogram of the individual noise terms. Apart from a high peak at
$\eta\approx 0.5625$ (which is just the maximum allowed value
$9/16$ within our numerical accuracy)
the plot is almost flat. The average is $\langle \, \eta
\, \rangle \approx 0$.

\begin{figure}[htb]
\includegraphics[width=7cm]{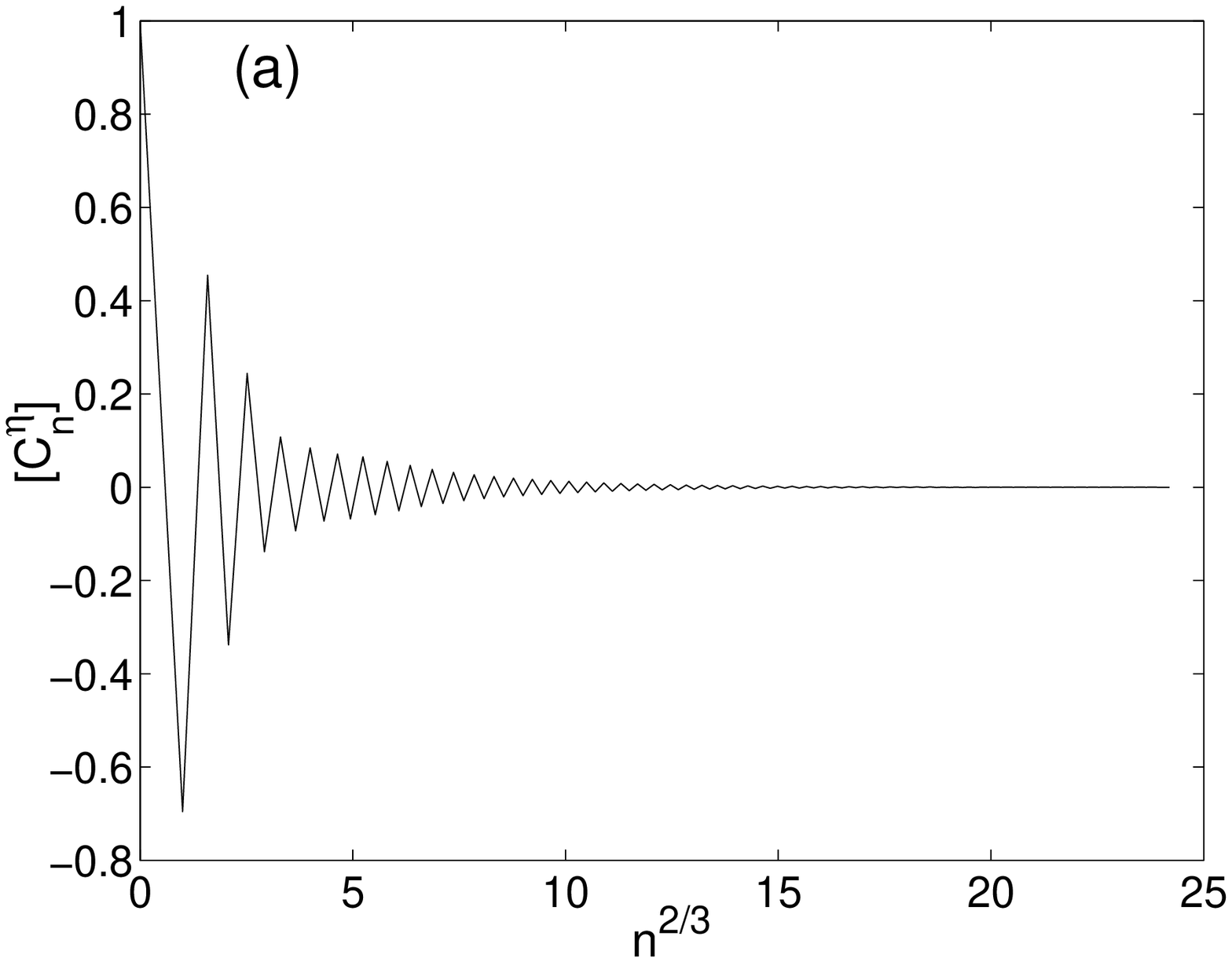}
\includegraphics[width=7cm]{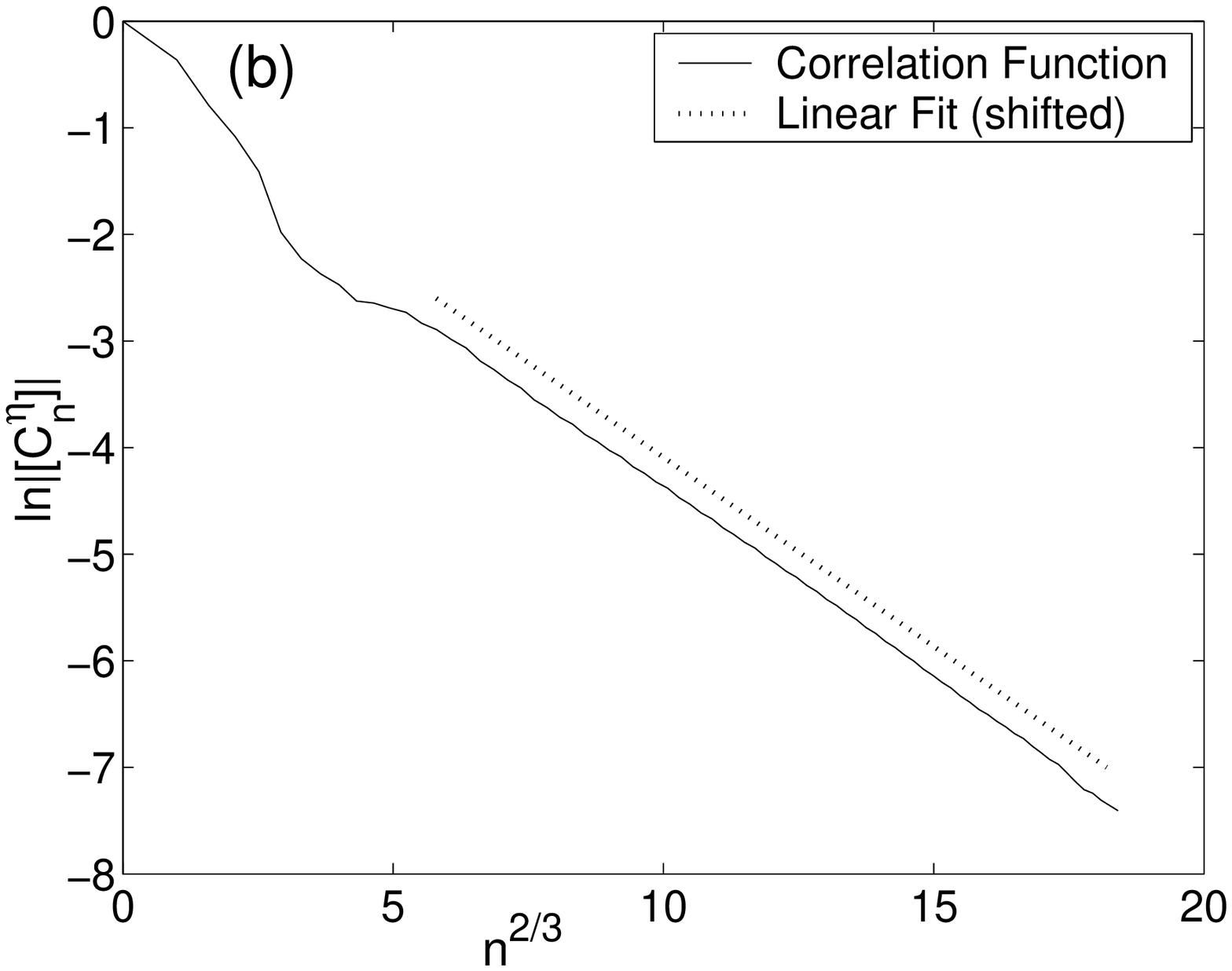}
\caption{Decay in time of the auto-correlation of the noise on
a site averaged over all sites in the sample, $[ \, C_n^\eta \, ]$,
as a function of $n^{2/3}$ [(a) linear plot, (b)
linear-logarithmic scale].}
\label{fig4}
\end{figure}

Since the connected correlation over time of a single map is rather
noisy we prefer to show its average over all sites in the system. More
precisely,  in Fig.~\ref{fig4}
we show the time decay of
\begin{equation}
[\, C^\eta_{n} \,]
\equiv
[\langle \, \Delta \eta_{k+n} \, \Delta \eta_{k} \, \rangle ]
=
\frac{1}{N} \sum_{i=1}^N
\left( \frac{1}{T-n} \sum_{k=0}^{T-n}
\eta^i_{k} \, \eta^i_{k+n}
\right)
\; ,
\end{equation}
where we used $\langle \, \eta_k \, \rangle \approx 0$.
Here and in what follows we use square brackets to denote an
average over space.
We  chose the scaling with $n^{2/3}$ of the
time axis and a logarithmic scale of the vertical axis to make
clear that the time-correlations decay as a stretched exponential:
\begin{equation}
[\, C^\eta_{n} \,]
\approx
e^{-\alpha n^{2/3}} \; .
\end{equation}
Even though we cannot exclude an additional power law dependence
on time this result demonstrates that the time correlations of the
noise $\eta$ are very short-ranged. Note the oscillations at short
times in Fig.~\ref{fig4}(a) that are not visible in Fig.~\ref{fig4}(b)
due to the absolute value.

\begin{figure}[htb]
\includegraphics[width=7cm]{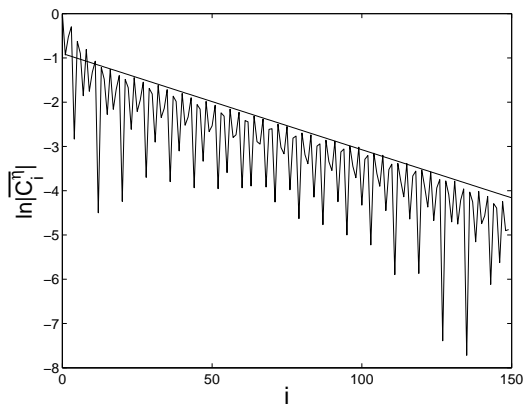}
\caption{Decay of the equal-time spatial correlations -
$\overline{C^\eta_i}$ defined in eq. (\ref{Cie}) (in
linear-logarithmic scale). The line corresponds to $e^{-i/\xi}$ with
$\xi \approx 46$.}
\label{other}
\end{figure}

We also studied the spatial correlations of $\eta$. In Fig.~\ref{other}
we display the equal time correlation
\begin{equation}
\overline{C^\eta_i} \equiv \frac{1}{T} \sum_{k=1}^T \left(
\frac{1}{N-i} \sum_{j=0}^{N-i} \eta^{j}_{k} \, \eta^{j+i}_{k}
\right) \; .
\label{Cie}
\end{equation}
Here and in what follows the overline indicates an average over time.
The line is a guide-to-the-eye showing the exponential decay
\begin{equation}
\overline{C^\eta_i} \approx e^{-i/\xi} \; , \qquad \qquad \xi\sim 46
\; ,
\end{equation}
defining a `correlation length', $\xi$, which is finite but rather long.
The oscillations can also be taken into account, but we omit
their precise functional description here.

\begin{figure}[htb]
\includegraphics[width=5cm]{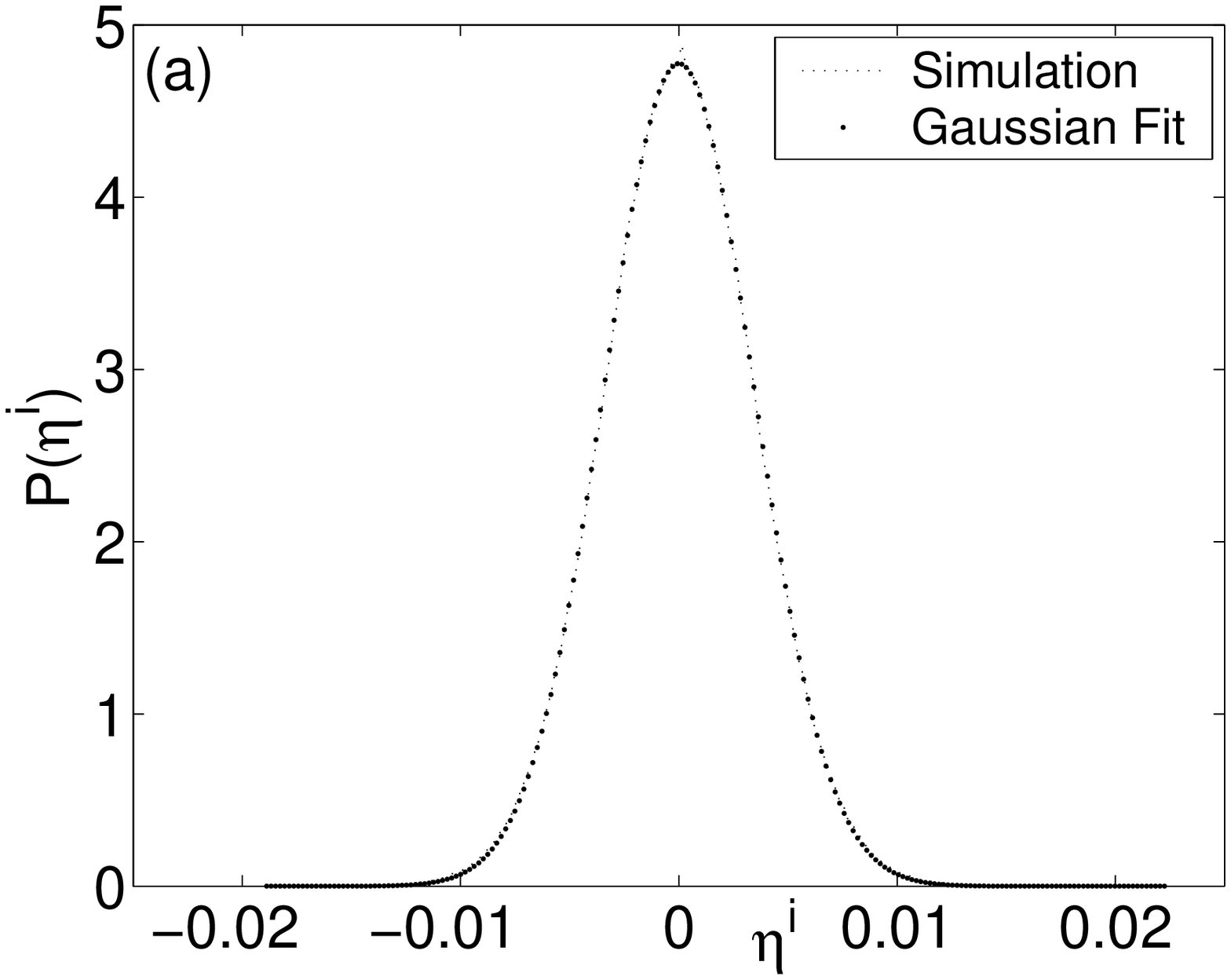}
\includegraphics[width=5.3cm]{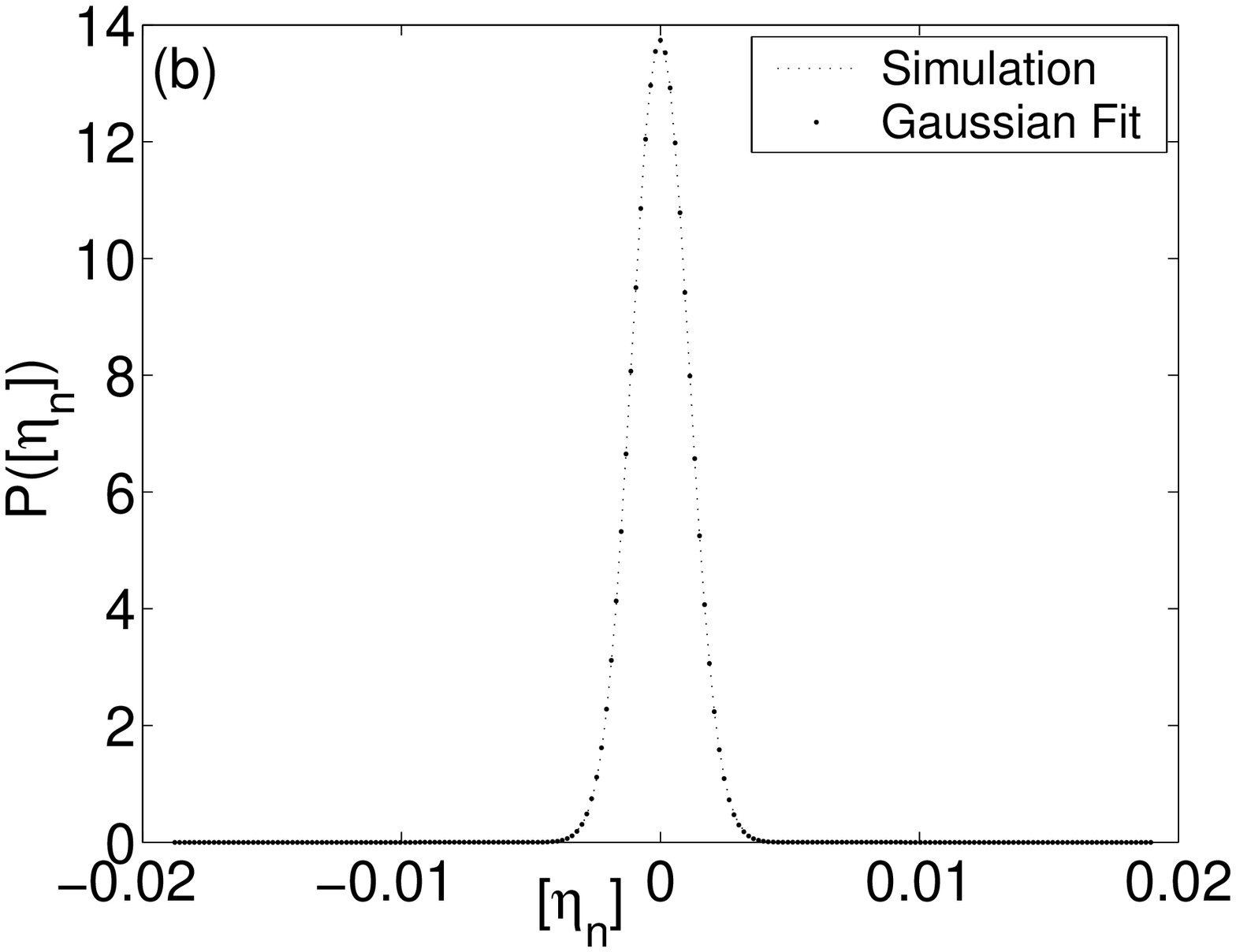}
\includegraphics[width=5.2cm]{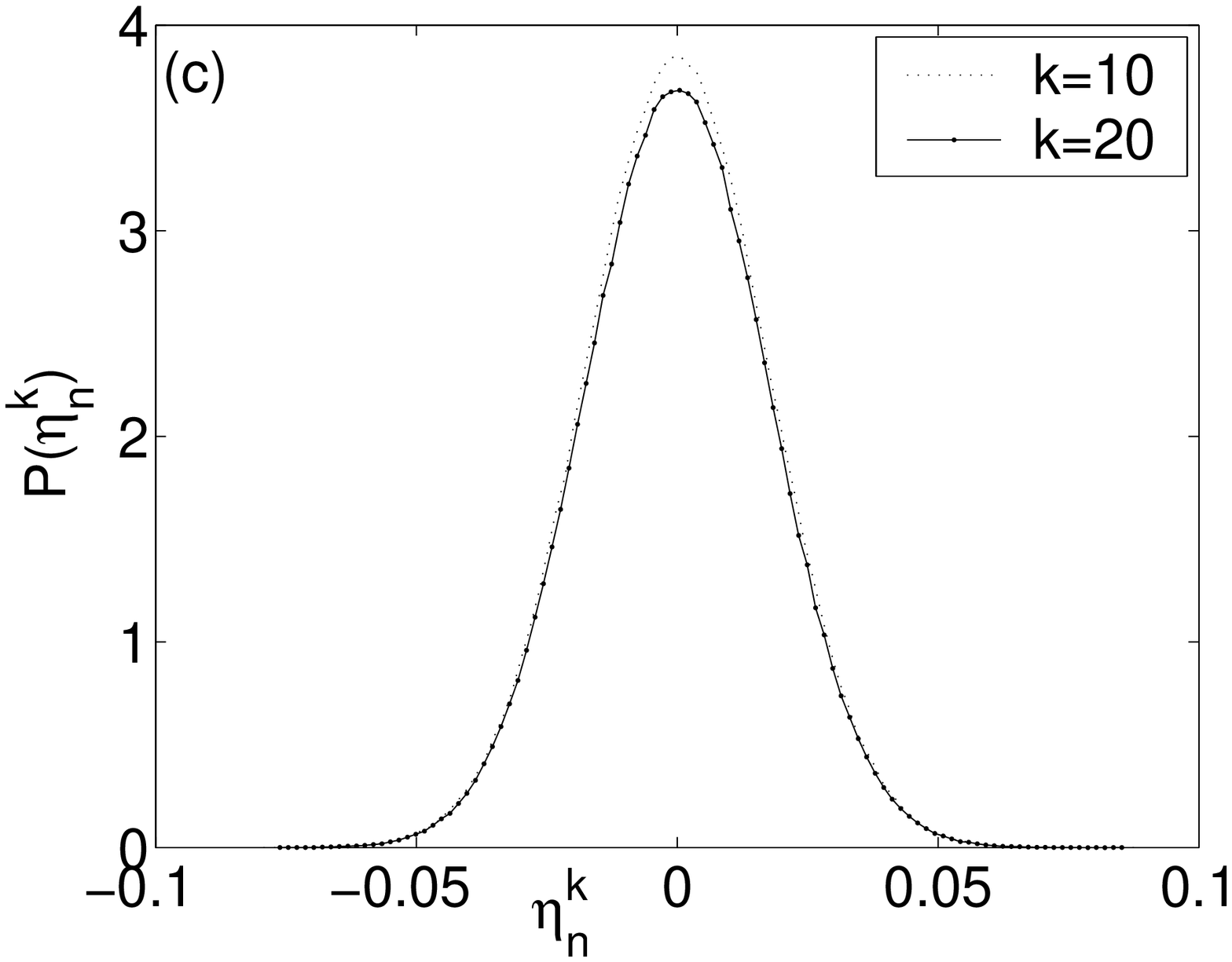}
\caption{The PDFs of  (a) $\overline{\eta^i}$, (b)
$[ \, \eta_n\,]$, and (c) Re$\left[\eta^k_n \right]$ for $k=10, \, 20$.}
\label{fig5}
\end{figure}

The short-range correlations in time and space shown in Figs.~\ref{fig4} and
\ref{other} are confirmed by an analysis of the PDFs of space and time
averages of the noise itself,
\begin{equation}
\overline{\eta^i} \equiv \frac{1}{T} \sum_{n=1}^T \eta_n^i
\; ,
\qquad \qquad
[ \, \eta_n \, ]  \equiv \frac{1}{N} \sum_{i=1}^N \eta_n^i
\; ,
\end{equation}
that are quite Gaussian, as shown in Fig.~\ref{fig5}(a) and
Fig.~\ref{fig5}(b), respectively. The data used to draw the histograms are
taken from the horizontal average at fixed vertical position, and
the vertical average at fixed horizontal position, respectively,
of the original continuous value of $x_n^i$ that gave rise to the
space-time map of Fig.~\ref{figa}(a).
One can also draw the histograms of different Fourier
components $\overline{\eta^k}$ [the case $k=0$ actually
corresponds to Fig.~\ref{fig5}(a)] which turn out to be also
Gaussian, as can be seen in Fig \ref{fig5}(c) where the cases
$k=10$ and $k=20$ are displayed.

\subsubsection{The confining aspect.}

In the previous subsection we studied the statistical properties of
the values taken by $\eta_n^i$ along a dynamical run.
We here analyse the effect of $\eta_n^i$ as a deterministic
force deriving from the potential (\ref{eq:Vh}).

\begin{figure}[htb]
\includegraphics[width=7cm]{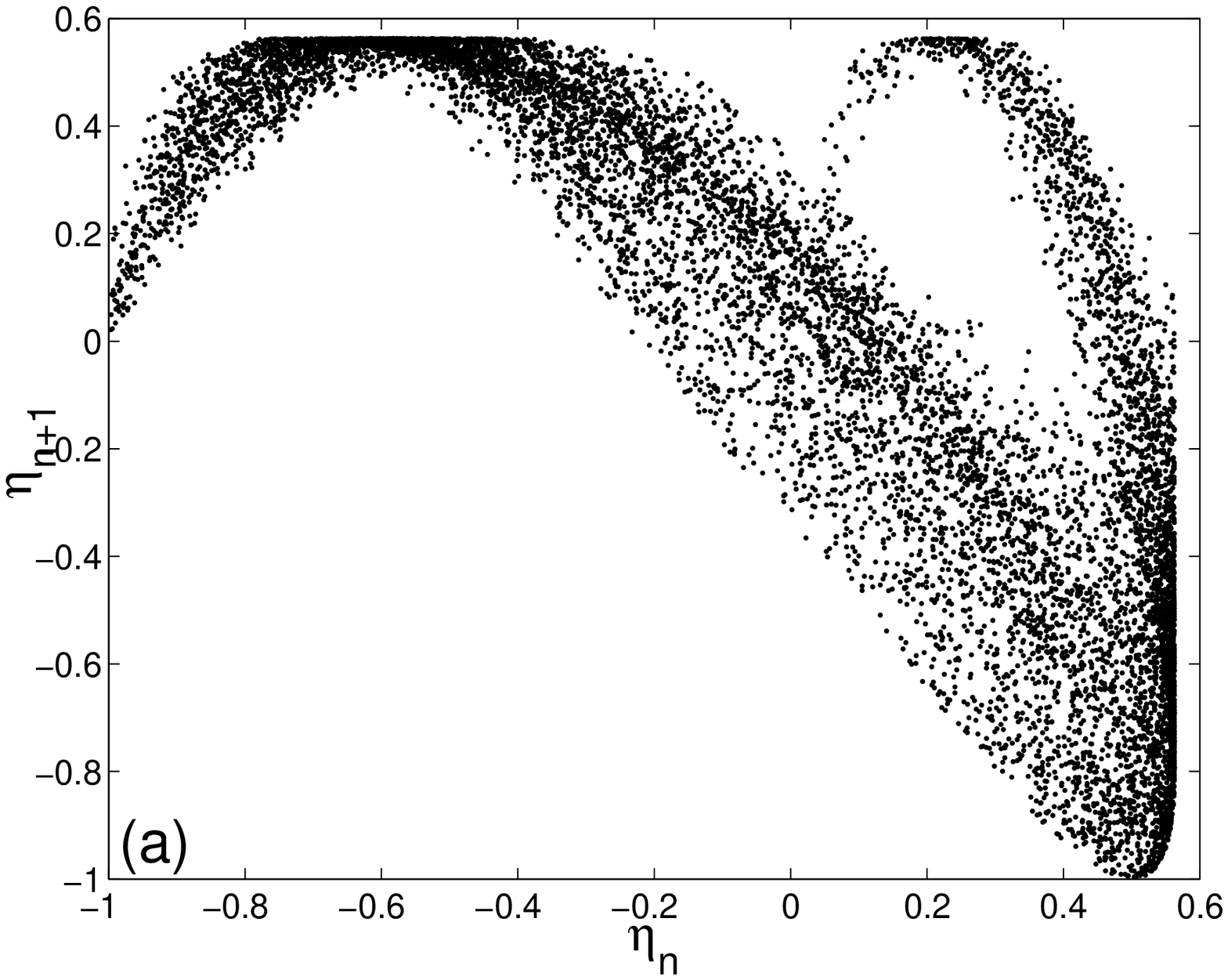}
\includegraphics[width=7.2cm]{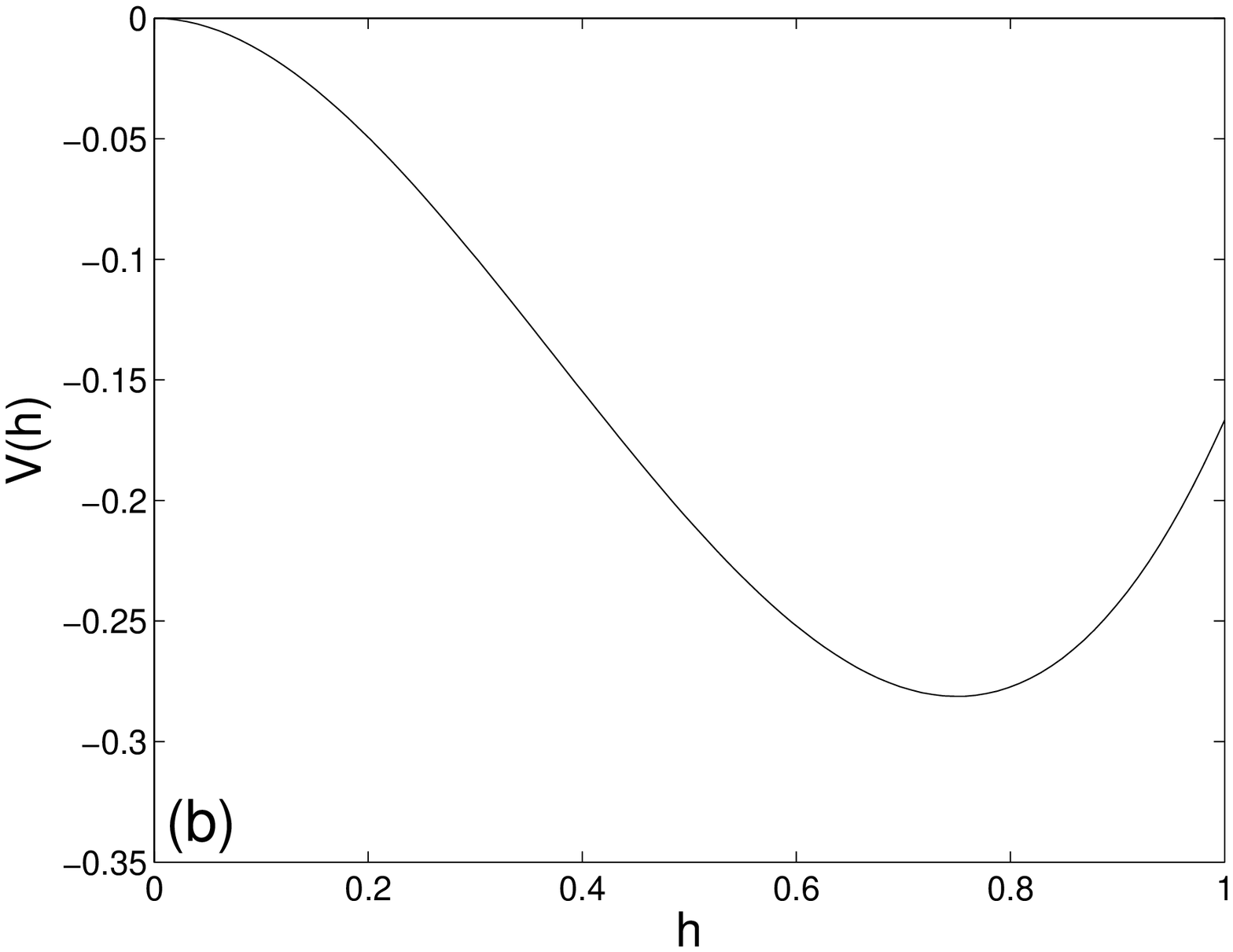}
\caption{(a) First return map, $\eta_{n+1}^j$ against $\eta_n^j$,
for a typical site in the fully chaotic regime ($r=4$). (b) The
confining potential $V(h)$ given in eq. (\ref{eq:Vh}).}
\label{return}
\end{figure}

The fact that $\eta_n^i$ is not really a perfect noise can be seen from
the study of the return maps.
Figure~\ref{return}(a) shows the first return map, {\it i.e.}
the plot of $\eta^j_{n+1}$ as a function of $\eta^j_n$ for many values
of $n$ and a chosen site on the lattice.
The information encoded in this plot is not included
in the analysis of the temporal correlations.
Even though the map is not a simple one-dimensional line, but
rather occupies some non-vanishing area, it does not fill phase
space. In other words, the return map is not structureless and
therefore in some sense not completely chaotic/stochastic (see
ref. \cite{Gonzalez}). This plot demonstrates that
unpredictability ({\it i.e.} no correlations) does not imply
randomness.

The bound on the original logistic map variables should translate
onto a confinement of the surface fluctuations $h$. The
interpretation of $\eta$ as a force deriving from the potential
$V(h)$ shown in Fig.~\ref{return}(b) allows one to identify
this term as an important one providing the confinement of the
fluctuations. For small $h$, more precisely, $h<(r-1)/r$ ($=3/4$
for $r=4$) the force is positive and it tends to increase the
value of $h$. Instead, for large $h$, $h>(r-1)/r$ the force is
negative and it pushes the fluctuation to take smaller values [a
similar confining effect is produced in the KS equation
(\ref{eq:KS})  by the term $-\kappa h$].

\subsubsection{The non-linear aspect.}

$\eta$ has the same structure as the non-linear terms in the
diffusive and otherwise linear FKPP equation. It is then clear
that eq.~(\ref{eq:CML}) admits the same spatially uniform and constant
fixed points $h=0$ and $h=(r-1)/r$. One can then expect to find
travelling wave solutions when special initial conditions are
chosen. We present some examples in Sect.~\ref{sec:travelling}
but we  delay their detailed numerical study to~\cite{prep}.

\subsection{The CML against KPZ: observables}
\label{sec:confront}

We here translate the definition of the observables of
interest in the KPZ framework to that of the CML and we confront
their behaviour in the two system.

\subsubsection{Confinement.}
\begin{figure}[htb]
\includegraphics[width=5cm]{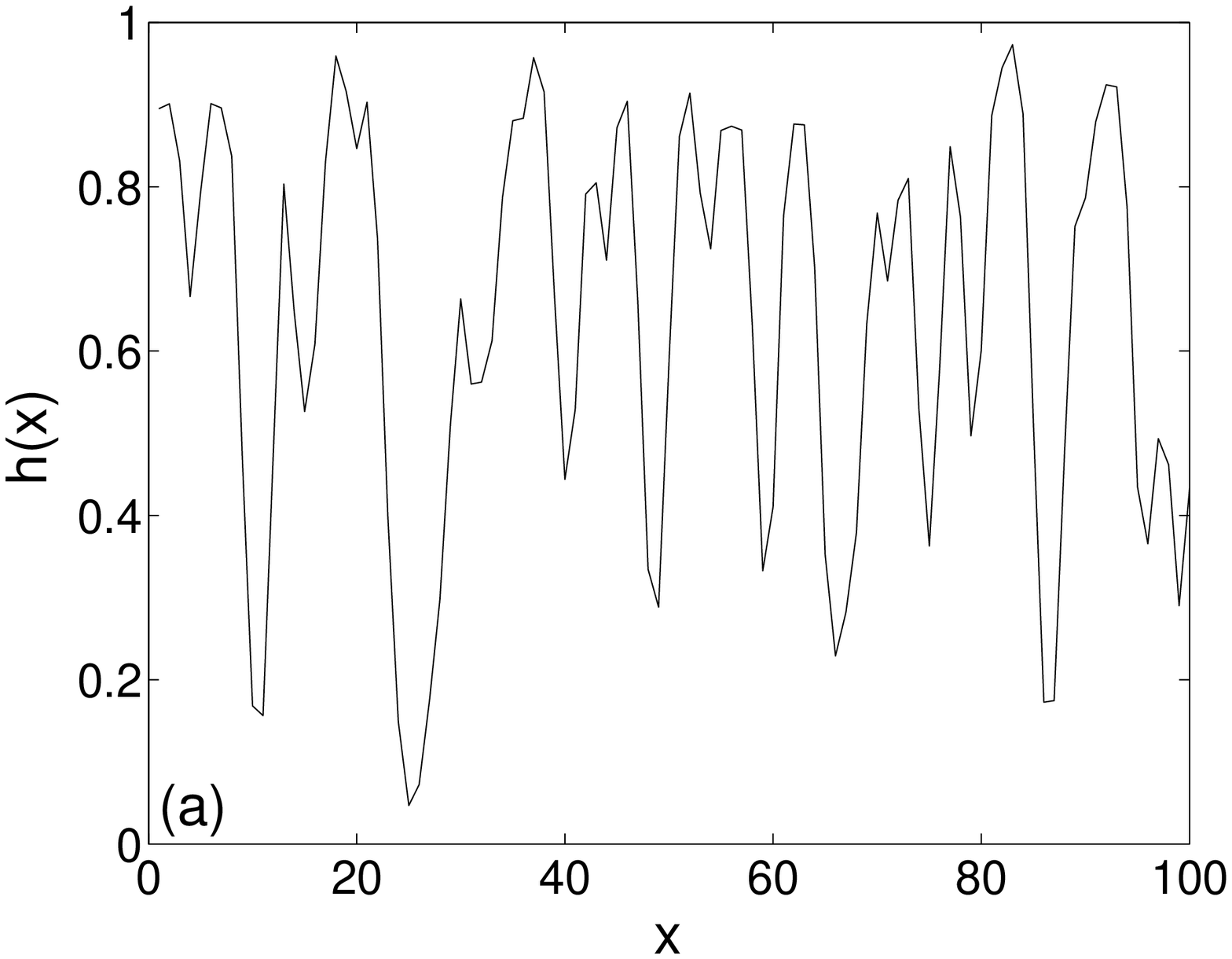}
\includegraphics[width=5cm]{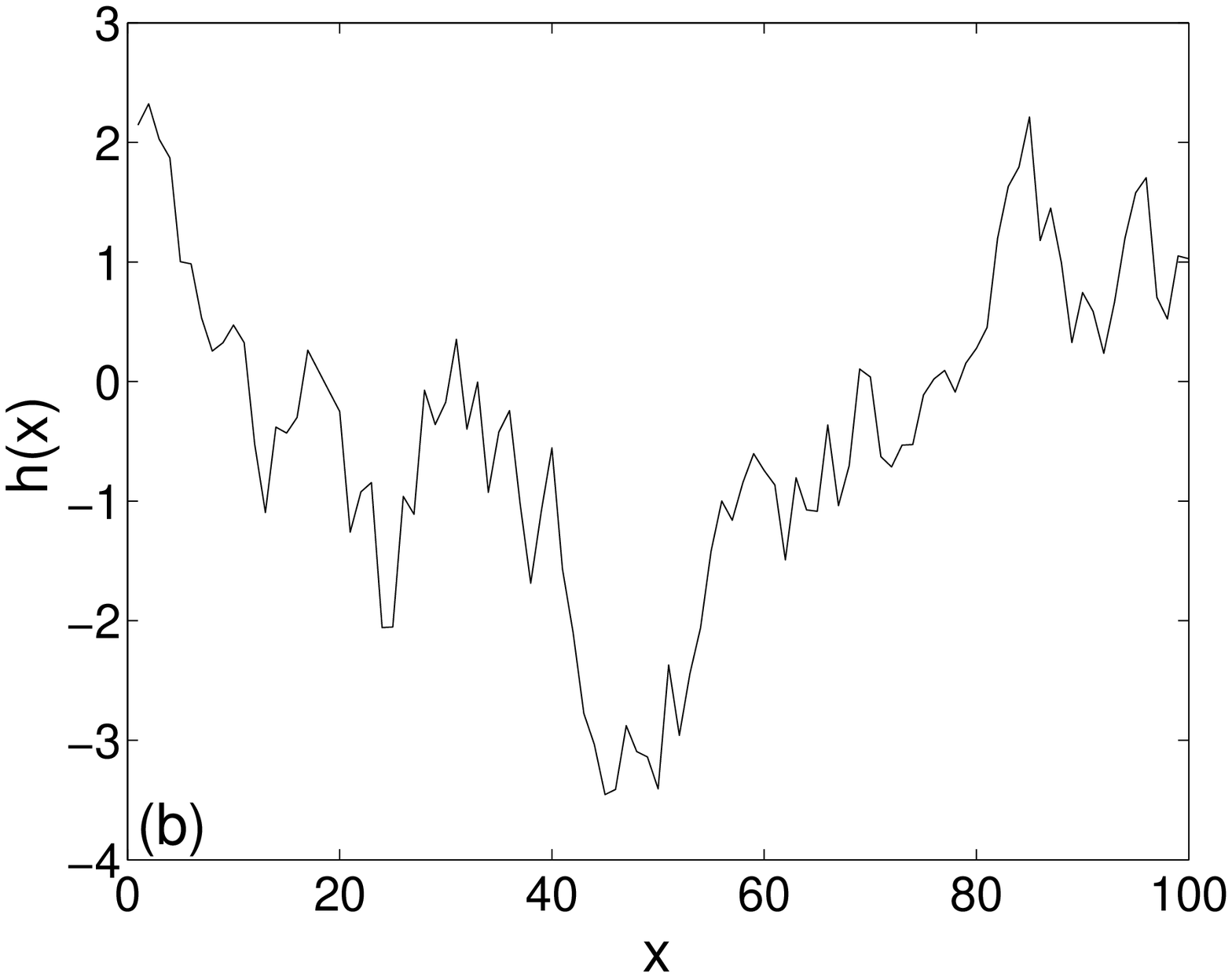}
\includegraphics[width=5cm]{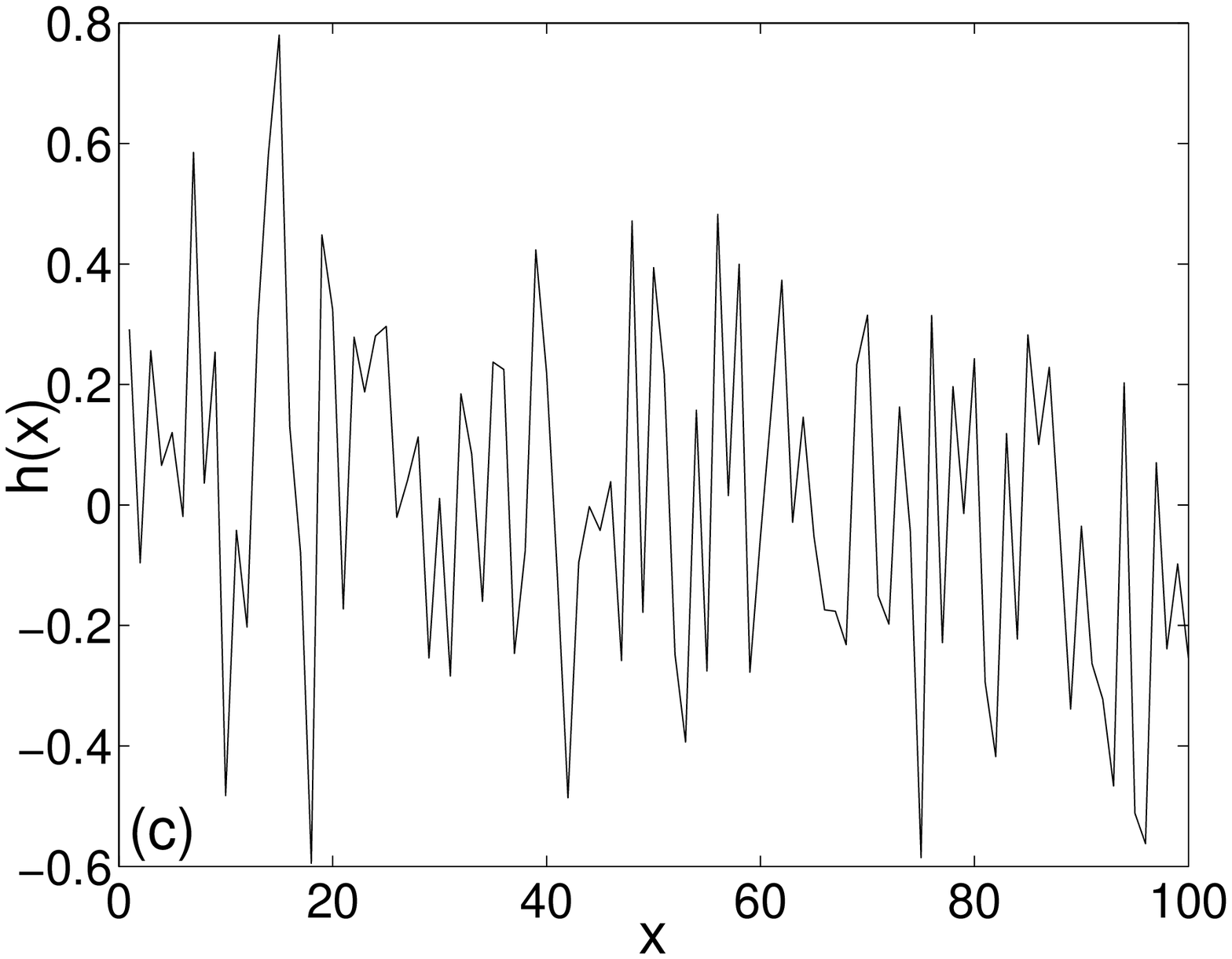}
\caption{(a) A CML interface, (b) a KPZ interface and (c) a
confined KPZ interface (with a confining potential of
$-\frac{1}{20}h^2$).}
\label{Interfaces}
\end{figure}

In the CML the variables are bounded and the mean-square
displacement cannot be larger than $1$. In Fig.~\ref{Interfaces}
we show a snapshot of a CML (a), a KPZ surface (b)
and a confined KPZ surface (c). At face value
these figures look different, although one can find similarities
between the CML and confined KPZ surfaces. Also, these figures are
very different from the KPZ surface being confined between two
`soft' walls~\cite{Munoz}.

Let us now compare the space and time dependence of these surfaces
in more detail.

\subsubsection{Roughness.}

Stochastic
surface growth from a flat initial configuration
is usually characterised in terms of the
mean-square displacement $w^2$.
In a one dimensional discrete model $w^2$ reads
\begin{equation}
w_\eta^2(L,t) \equiv \frac{1}{L} \sum_{i=1}^L
\left[h^i(t) - \frac1L \sum_{j=1}^L h^j(t) \right]^2
\; .
\end{equation}
The square root of the noise averaged displacement,
$w(L,t) \equiv \sqrt{\langle \, w_\eta^2(L,t) \, \rangle }$
has the scaling behaviour~\cite{Barabasi-Stanley}
\begin{eqnarray}
w(L,t) \sim L^{\alpha}
f\left( \frac{t}{L^{z}} \right)
\; ,
\qquad
f(u) \sim
\left\{
\begin{array}{ll}
u^{\frac{\alpha}{z}} \;, \; & u\ll 1 \;\; (\mbox{growth})
\; ,
\\
{\mbox const} \;, \; & u\gg 1 \;\; (\mbox{saturation}) \; ,
\end{array}
\right.
\label{Family}
\end{eqnarray}
with $\alpha$ and $z$ the roughness and dynamic exponents,
respectively.

\begin{figure}[htb]
\includegraphics[width=7cm]{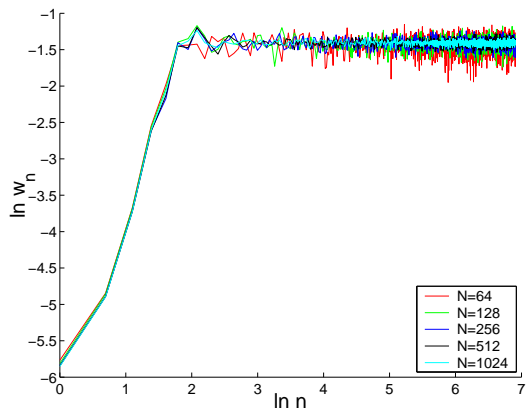}
\caption{Roughening of a CML with small random initial conditions
for various system sizes given in the key.}
\label{Roughening}
\end{figure}

Since the CML surface is necessarily bounded,
if the scaling form (\ref{Family}) holds, $\alpha$ must be identically
zero.
If we wish to check $z$ numerically we need to specify the initial
conditions. In the CML a flat configuration $x_0^i=x$ for all
sites remains the same at all later times. In order to generate a
rough configuration we need to introduce some disorder initially.
This can be done in a number of ways. For example, one can analyse
the motion of an initial localised bump on an otherwise flat
configuration or one can add some small random noise on a flat
state. In Fig.~\ref{Roughening} we present results of a
simulation of the CML, with random, though very small, initial
condition and for several system sizes.

\noindent As can be seen in the figure, the saturation time does
not depend on the size of the system, and occurs after $\sim 10$
time steps in all cases. Also, the width in the steady-state ({\it
i.e.} $w_n$ for large $n$'s) is also independent of system size.
These findings are consistent with vanishing exponents $z=0$ and
$\alpha=0$ [see eq.~(\ref{Family})]. Actually, vanishing $\alpha$
and $z$ is also consistent with a simple scaling argument applied
to the confined KPZ equation, with a confining potential of
$-\frac{\kappa}{2}h^2$ (at this point it is worthwhile mentioning
that some confined KPZ systems have been studied in the
past~\cite{Munoz} though,
as far as we know, non is directly relevant here).

The non-trivial character of the PDF of mean-square displacements
found in several unbounded surface growth problems~\cite{Racz}
is completely erased by the confinement.

\subsubsection{Stretched exponential relaxation.}

\begin{figure}[htb]
\includegraphics[width=7cm]{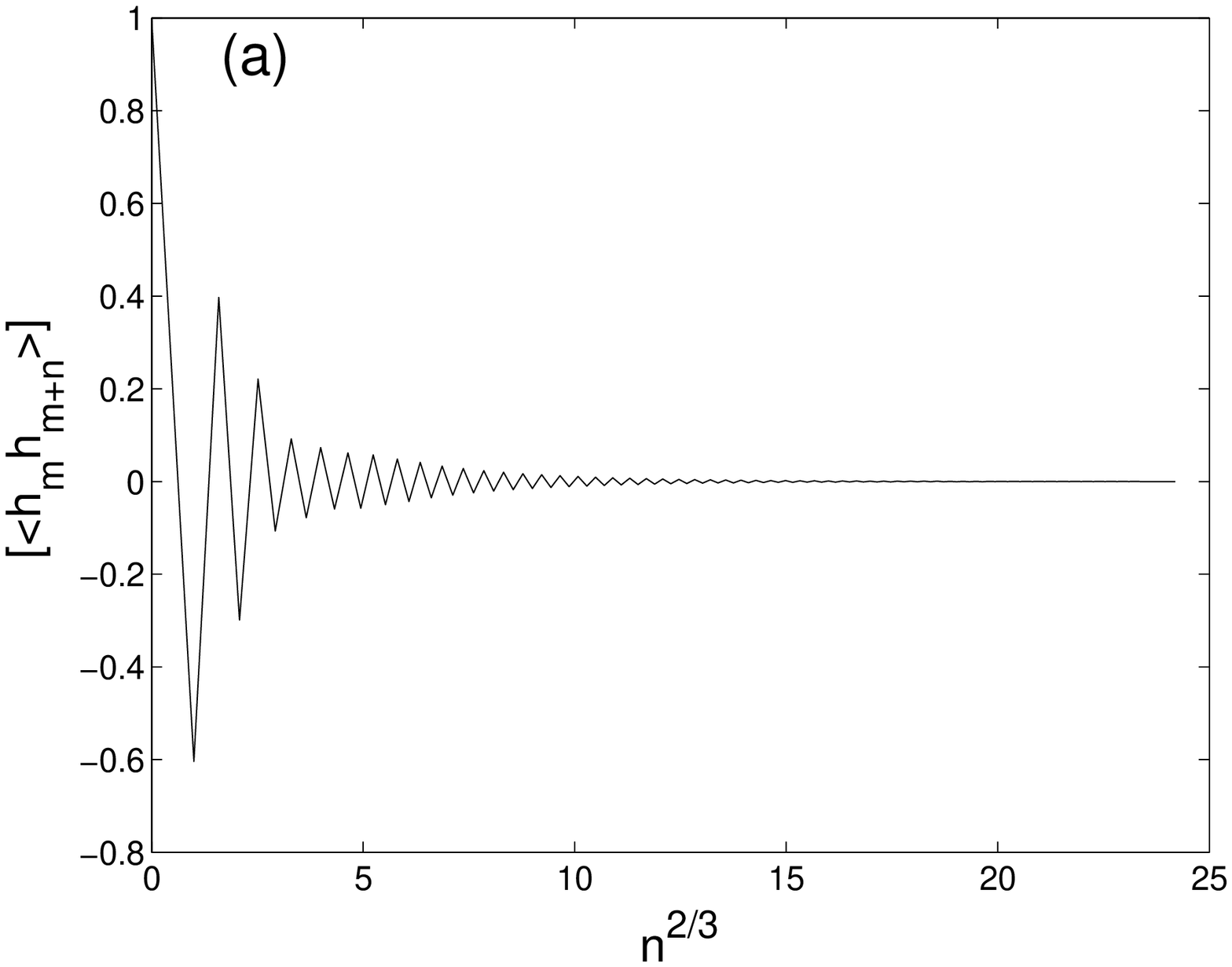}
\includegraphics[width=6.8cm]{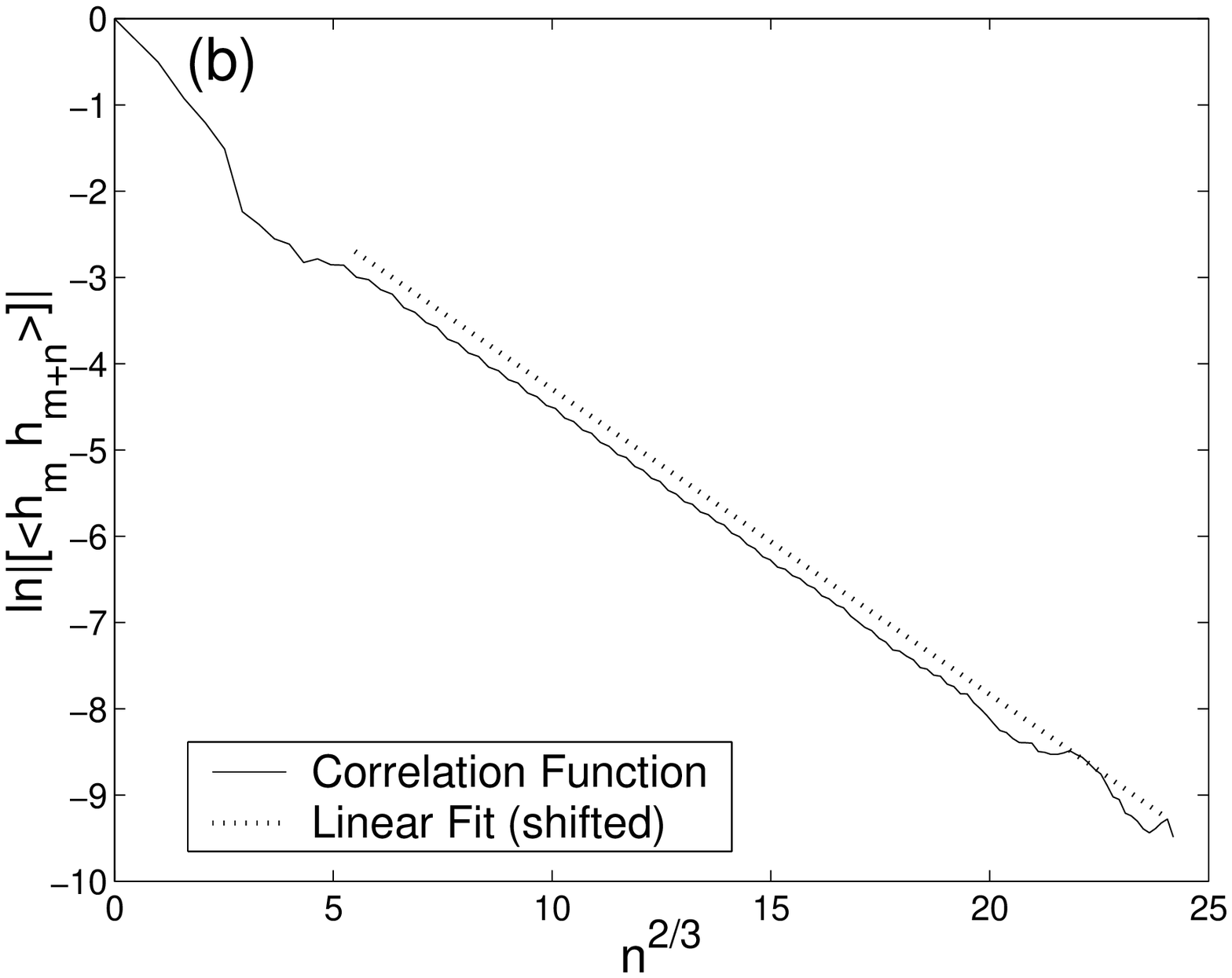}
\caption{The correlation function $[\, \langle \, h_{m+n} \, h_{m}
\, \rangle ]$ as a function of $n^{2/3}$ [(a) linear plot,
(b) linear-logarithmic scale).}
\label{CMLcor}
\end{figure}

The decay in time of the field-field correlations is still
non-trivial in this system and very similar to the one found for
the (usual and unbounded)  KPZ equation. In
Fig.~\ref{CMLcor} we show the space averaged, time correlation of
the field $[\, \langle \, h_{m+n} \, h_{m} \, \rangle ]$.
The linear dependence in the linear-logarithmic
scale used in panel (b) indicates the stretched exponential
decay
\begin{equation}
C_n^h(t) \approx e^{-n^\zeta}
\; ,
\qquad \mbox{with}
\qquad
\zeta=2/3
\; ,
\end{equation}
The panel (a) shows that the decay occurs in an oscillatory
way.

A stretched exponential relaxation of the dynamical structure
factor in the $d=1$-KPZ growth,
\begin{equation}
C(k, t) \equiv
\langle \, h(k, t) h^*(k,0) \, \rangle
\sim A e^{-B k t^{1/z}}
\end{equation}
with $z$ the dynamic exponent was found by solving the KPZ equation
within the mode-coupling approximation~\cite{Moore}, with a
self-consistent expansion~\cite{Moshe} and with a direct numerical
integration in $d=1$~\cite{Eytan}.  An oscillatory superimposed
dependence on time was obtained numerically~\cite{Moore,Eytan} and can
also be obtained analytically from the self-consistent
expansion~\cite{Eytan2}.  Some doubt on this result was shed by the
solution to a polynuclear growth problem that is in the same
universality class as $d=1$ KPZ~\cite{Spohn} with a dynamic structure
factor with a simple exponential decay.  This discrepancy may be due
to the fact that defining universality classes out of equilibrium
might tricky: some models may share the static exponents but differ in
their overall dynamical behaviour~\cite{Eytan}.

\begin{figure}[htb]
\includegraphics[width=5cm]{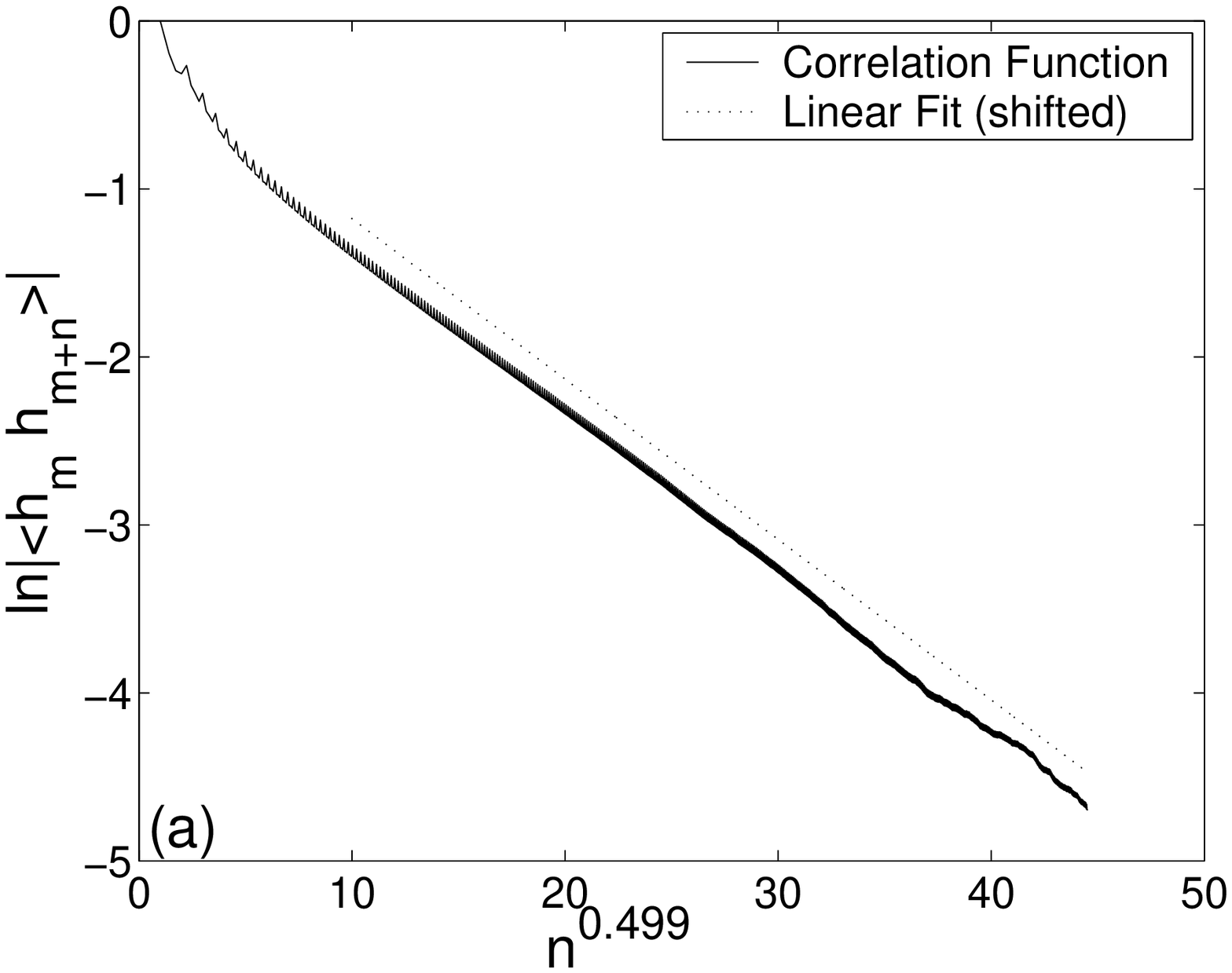}
\includegraphics[width=5cm]{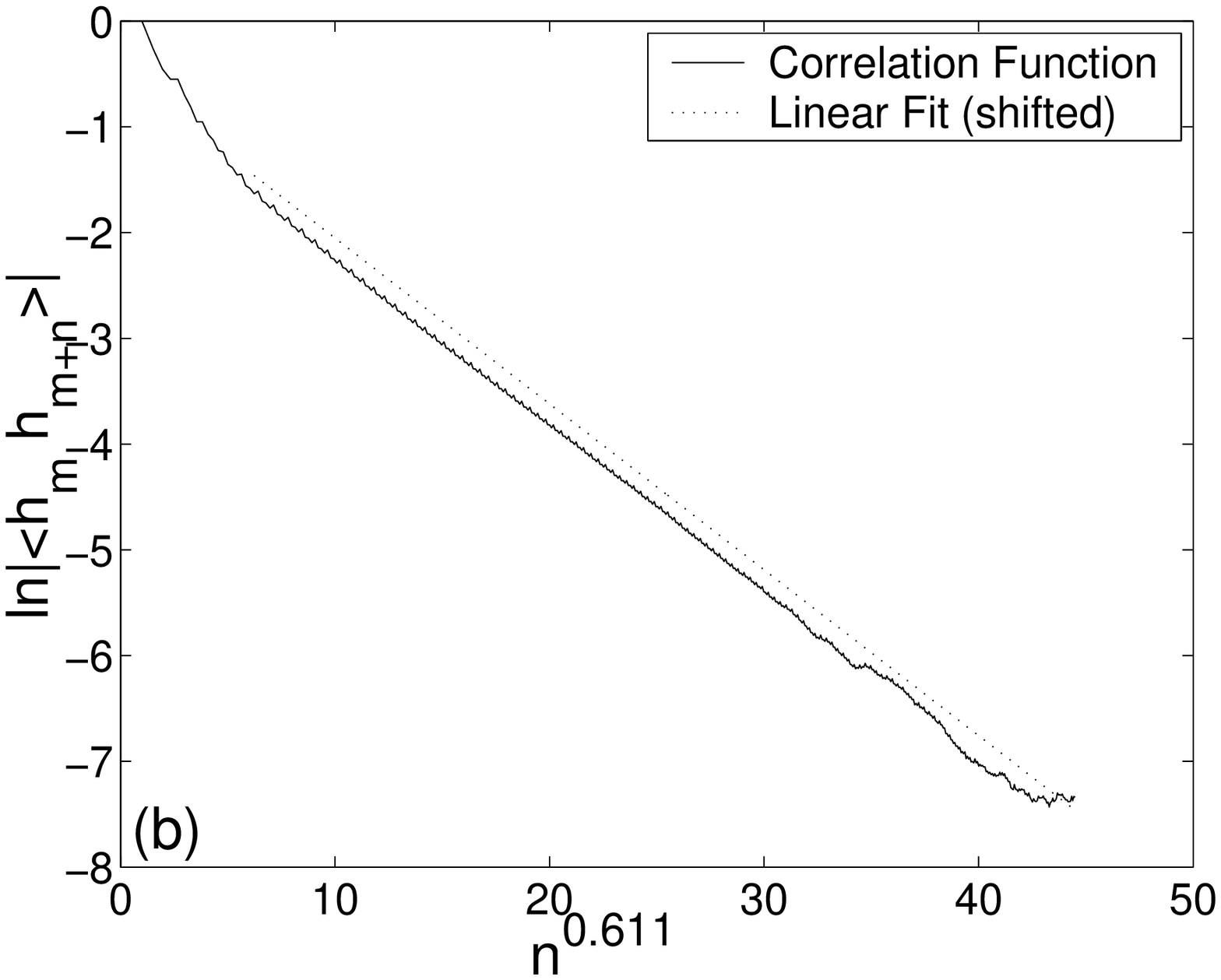}
\includegraphics[width=5cm]{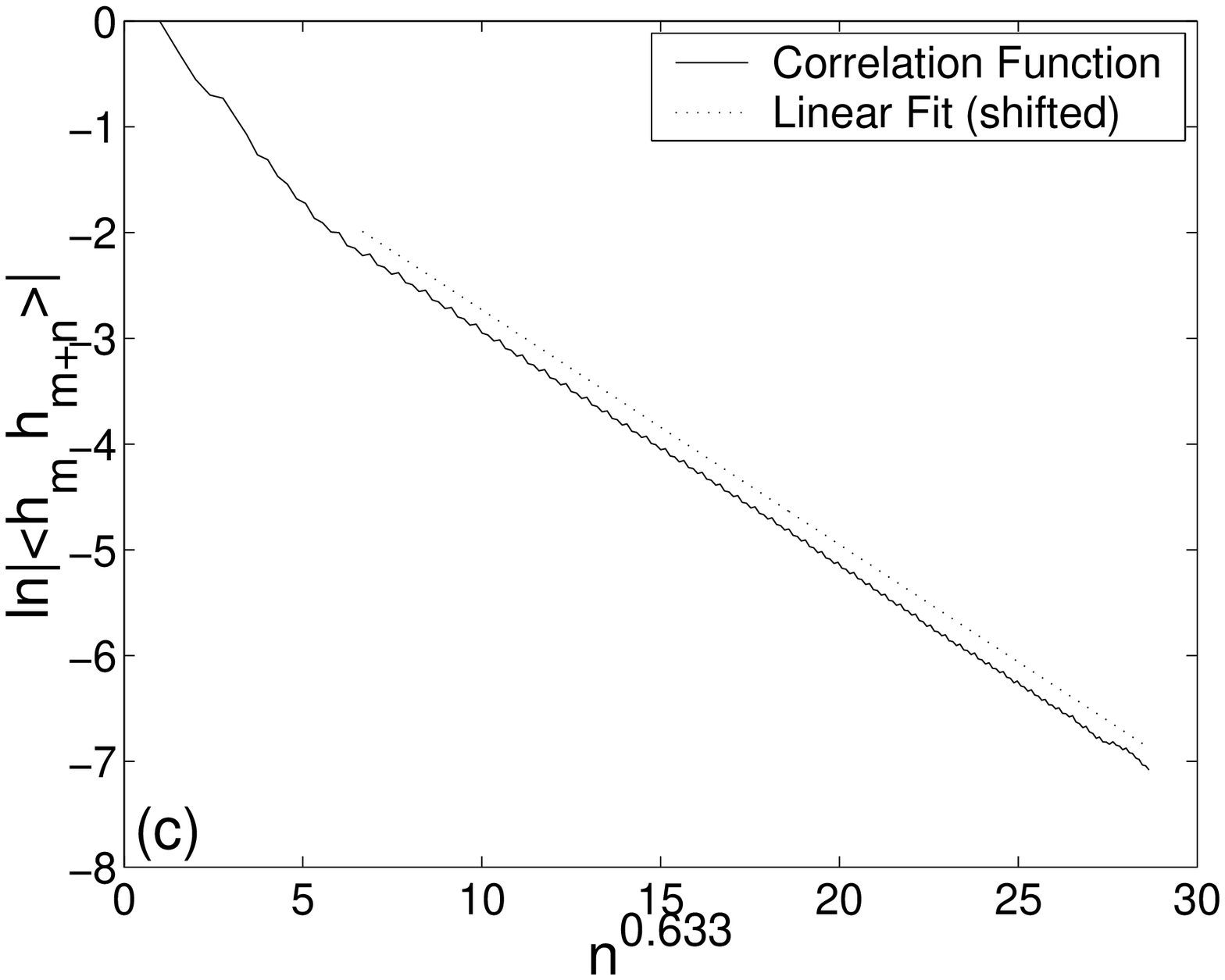}
\caption{The $r$ dependence of the stretched exponential relation
for (a) $r=3.83$, (b) $r=3.87$ and (c) $r=3.9$.}
\label{r-dependence}
\end{figure}

We thus found that the dynamic structure factor of the CML of
logistic units in their deep chaotic regime ($r=4$) and the $d=1$
KPZ equation have the same stretched exponential relaxation.
Surprisingly enough, the dynamic exponent $z=3/2$ is reproduced in
the CML model where it does not have the interpretation of
relating spatial and time fluctuations: the study of the
mean-square fluctuations yields a trivial size independent result
($z=0$) due to confinement. The dependence of the exponent $\zeta$
on the non-linear parameter $r$ is weak, as demonstrated in
Fig.~\ref{r-dependence} where we display the decay of the dynamic
structure factor at $k=0$ for several values of $r$.

In a sense, it is not surprising that the bound on the values of the
fluctuating field does not affect the time decay of the correlations.
In the context of glassy systems stretched exponential relaxations
of correlation functions were searched and found in kinetically constrained
models~\cite{Buhot} in which the dynamic variables are naturally bounded
(they are spins).

It would be very interesting to test whether a similar stretched exponential
relaxation occurs in the confined continuous KPZ equation but this
check goes beyond the scope of this article.

\subsection{The Lyapunov vector}
\label{sec:Lyapunov}

It was conjectured by Pikovsky {\it et al}~\cite{Politi} that the
evolution of linear perturbations in the chaotic regime of a CML
of logistic elements is well described by the KPZ stochastic
dynamics. The notion of a Lyapunov vector is one of the
ways to extend the notion of Lyapunov exponent to space-time chaos.
One solves simultaneously the original
nonlinear dynamical system, and the linearized equation for a
perturbation. The
Lyapunov exponent is then determined from the norm of the Lyapunov
vector.

In the case of the CML, it is convenient to write the linearisation of
eq.~(\ref{eq:coupled-maps}) in the following way
\begin{eqnarray}
\tilde \omega({x,t}) = f'\left[ {h({x,t})} \right]
\omega( {x,t})
\label{eq:linearized}
\; ,
\\
\omega({x,t + 1}) =
\left( {1 - \nu} \right)
\tilde\omega({x,t}) +
\frac{\nu}{2}
\left[ \tilde\omega({x - 1,t}) + \tilde \omega({x + 1,t}) \right]
\; ,
\label{eq:linearized1}
\end{eqnarray}
where $\omega(x,t)$ is the Lyapunov vector. using the continuum limit
(\ref{eq:continuum1})-(\ref{eq:continuum4}) we obtain the following
continuum equation for $\tilde \omega(x,t)$:
\begin{equation}
\frac{\partial\tilde\omega}{\partial t} =
\frac{\nu}{2}
\frac{\partial^2 \tilde\omega}{\partial x^2}
+
\left\{f'\left[h({x,t + 1}) \right] - 1 \right\}
\;
\left[
\tilde \omega({x,t}) + \frac{\nu}{2}
\frac{\partial^2 \tilde \omega}{\partial x^2}
\right]
\; .
\end{equation}
We are then led to identify a `multiplicative noise'
\begin{equation}
\xi _ \times  \left( {x,t} \right)
\equiv
f'\left[ {h\left( {x,t + 1} \right)} \right] - 1
\; .
\end{equation}
(The fact that $f'\left[ {h\left( {x,t + 1} \right)} \right]$ is
strictly speaking evaluated at time $t+1$ is not a real problem since the
process does not depend on the evolution of $\omega(x,t)$, as it
does not feed into the equation for the $h$'s. Therefore, one can
say that intrinsically the noise in our system should be
understood in the It\^o sense.)

At this point, since the variable $\tilde \omega(x,t)$ is not
bounded, a simple scaling argument can convince us that the
coupling of the second derivative to the noise $\xi_x$ is less
relevant (in the RG sense) than the coupling to $\tilde
\omega(x,t)$ itself. Therefore, we neglect the last term inside
the square brackets. The equation we are left with is just a
diffusion equation in the presence of multiplicative noise. This
motivates the application of the Hopf-Cole transformation, $\tilde
H = \ln \tilde \omega$, which yields
\begin{equation}
\frac{{\partial \tilde H}} {{\partial t}} = \frac{\nu }
{2}\frac{{\partial ^2 \tilde H}} {{\partial x^2 }} + \frac{\nu }
{2}\left( {\frac{{\partial \tilde H}} {{\partial x}}} \right)^2  +
\xi _ \times  \left( {x,t} \right) \label{KPZlike}.
\end{equation}

Now, we have to figure out the dynamics of $H \equiv \ln \omega$,
{\it i.e.} the Lyapunov vector. This is easily done by taking the
$\ln$ of eq.~(\ref{eq:linearized})
\begin{equation}
H\left( {x,t} \right) = \tilde H( {x,t}) - \ln
f'\left[ {h\left( {x,t} \right)} \right]
\end{equation}
This means that $H$ will be just $\tilde H$ up to an additional
`additive noise', $\xi _ +  \left( {x,t} \right)$, defined as
\begin{equation}
\xi _ +( {x,t}) \equiv   - \ln f'\left[ {h\left( {x,t}
\right)} \right]
\; .
\end{equation}
The roughness of the `interface' $H( {x,t})$ is
given by the equal-time autocorrelation function of
$H({x,t})$. Therefore, we need to check the statistical
properties of $\xi _ +( {x,t})$, that is to say,  its
autocorrelation function and its correlations with $\tilde
H({x,t})$. We checked that since the term is
exponentially distributed, with very short-range correlations in
space-time and vanishing correlations with $\tilde H({x,t})$,
the dominant contribution to the autocorrelation
function of $\tilde H( {x,t})$ comes from $H({x,t})$,
and is therefore described by the KPZ
exponents. This is not surprising, since the noise $\xi_+(x,t)$
does not accumulate in $H(x,t)$, but rather it just adds a random
number to $\tilde H(x,t)$ with a small amplitude.

This establishes the hypothesis of Pikovsky
{\it et al}~\cite{Politi} but, more importantly, it allows us to make a step
forward relating known quantities in the KPZ literature to
features of the Lyapunov exponent. The Lyapunov
exponent is obtained from the norm of the Lyapunov vector. If one uses
the so-called $0$-norm,
\begin{equation}
N_0(t)=\exp \left[\frac{1}{L}\int_0^L h(x,t)dx\right]
\; ,
\end{equation}
it is then assured to be a self-averaging quantity. In addition,
the Lyapunov exponent is given by
\begin{equation}
\lambda  = \mathop {\lim }\limits_{T \to \infty } \frac{{\ln N_0
\left( T \right) - \ln N_0 \left( 0 \right)}} {T}
\; ,
\end{equation}
and this is no other than the large-deviation function for the
Asymmetric Exclusion Process (ASEP).  calculated previously by Derrida
and Appert \cite{Appert}.  The ASEP is a discrete model in the
universality class as KPZ in one dimension.  In terms of the ASEP,
the Lyapunov exponent is given by
\begin{equation}
\lambda  = \rho \left( {1 - \rho } \right) + \sqrt {\frac{{\rho
\left( {1 - \rho } \right)}} {{2\pi L^3 }}} G\left( {\sqrt {2\pi
\rho \left( {1 - \rho } \right)L} } \right) \label{lyapunov}
\label{eq:lambda-DA}
\end{equation}
where $L$ is the system size, $\rho$ is a the density of particles
(a parameter in ASEP), and $G(\beta)$ is a scaling function
independent of $L$ and $\rho$ and known is an implicit form
(see~\cite{Appert} for more details).

What is left now is to relate the parameters of the ASEP model to
those of the the KPZ equation we derived for $H(x,t)$
(\ref{KPZlike}). Using ref. \cite{ASEP} we see that
$\rho(1-\rho)=D/32$ where $D$ is the noise amplitude characterizing
$\xi_\times(x,t)$. In general, $D$ would be a complicated function of
$r$ and $\nu$, and is unknown at the moment. Still, the expression for
the Lyapunov exponent (\ref{lyapunov}) could be useful to study finite
size effects. As an exercise, we try to estimate $D$ using the known
PDF of a single (uncoupled) map, in the fully chaotic regime ($r=4$),
given by eq.~(\ref{eq:px}). In that limit $\xi_\times = 3-8x$, and is
known to be uncorrelated in time. Therefore,
$D=\langle\xi_\times^2\rangle=9$. Plugging this into
eq.~(\ref{eq:lambda-DA}) yields $\lambda=9/32 \simeq 0.28$. In
numerical simulations Pikovsky and Politi found $\lambda \simeq
0.38$~\cite{Politi} which is of the same order of magnitude as our
result. The difference is certainly related to the fact that we
approximated $\xi_\times$ by that of a single map. In addition, we
neglected $\xi_+$, and this could have influenced as well.

\section{Travelling wave solutions}
\label{sec:travelling}

Let us now briefly analyse the coupled map lattice model  and its
continuum limit from the viewpoint of front propagation and look for
travelling wave configurations linking the two stationary solutions:
\begin{equation}
h=0 \;, \qquad \mbox{and} \; \qquad h=\frac{r-1}{r}
\; .
\end{equation}
The non-linear term $\eta$ renders the solution $h=0$ linearly unstable:
$d\eta/dh(h=0)= (r-1)>0$ and the solution $h=(r-1)/r$ stable
$d\eta/dh(h=(r-1)/r) <0$ if $r>1/2$.
Introducing the travelling wave {\it Ansatz} (\ref{eq:travel}) in
the linearized version of eq.~(\ref{eq:CML})
in which we {\it drop} the terms
proportional to $F^2$, $F F''$ and $(F')^2$ we find
\begin{equation}
v \geq v_{min} \qquad \mbox{with} \qquad
v_{min} = \sqrt{2 \nu r(r-1)}
\label{eq:v-min}
\; ,
\end{equation}
and $\gamma_{min} = \sqrt{2(r-1)/(\nu r)}$.
The combined effect of the diffusion and non-linear
term of KPZ type are not obvious but one may expect that
if we start with a localized initial condition
with, {\it e.g.}, the form of a bump with $h\neq 0$ in a finite
region of the one dimensional space, the non-linear term will drive
the border of the bump into the unstable state $h=0$. The front may
then advance as a travelling wave.

\begin{figure}[htb]
\includegraphics[width=7cm]{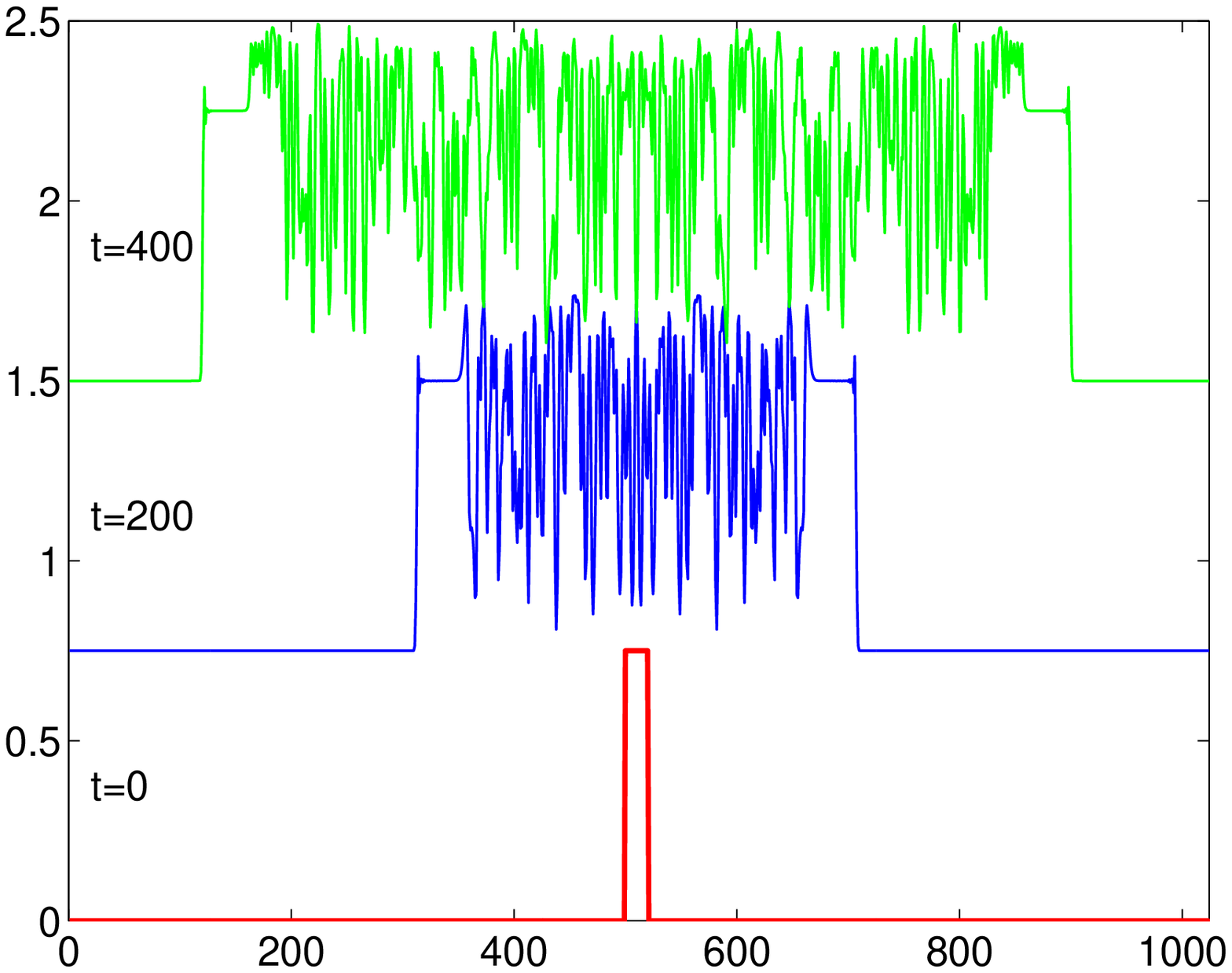}
\includegraphics[width=7cm]{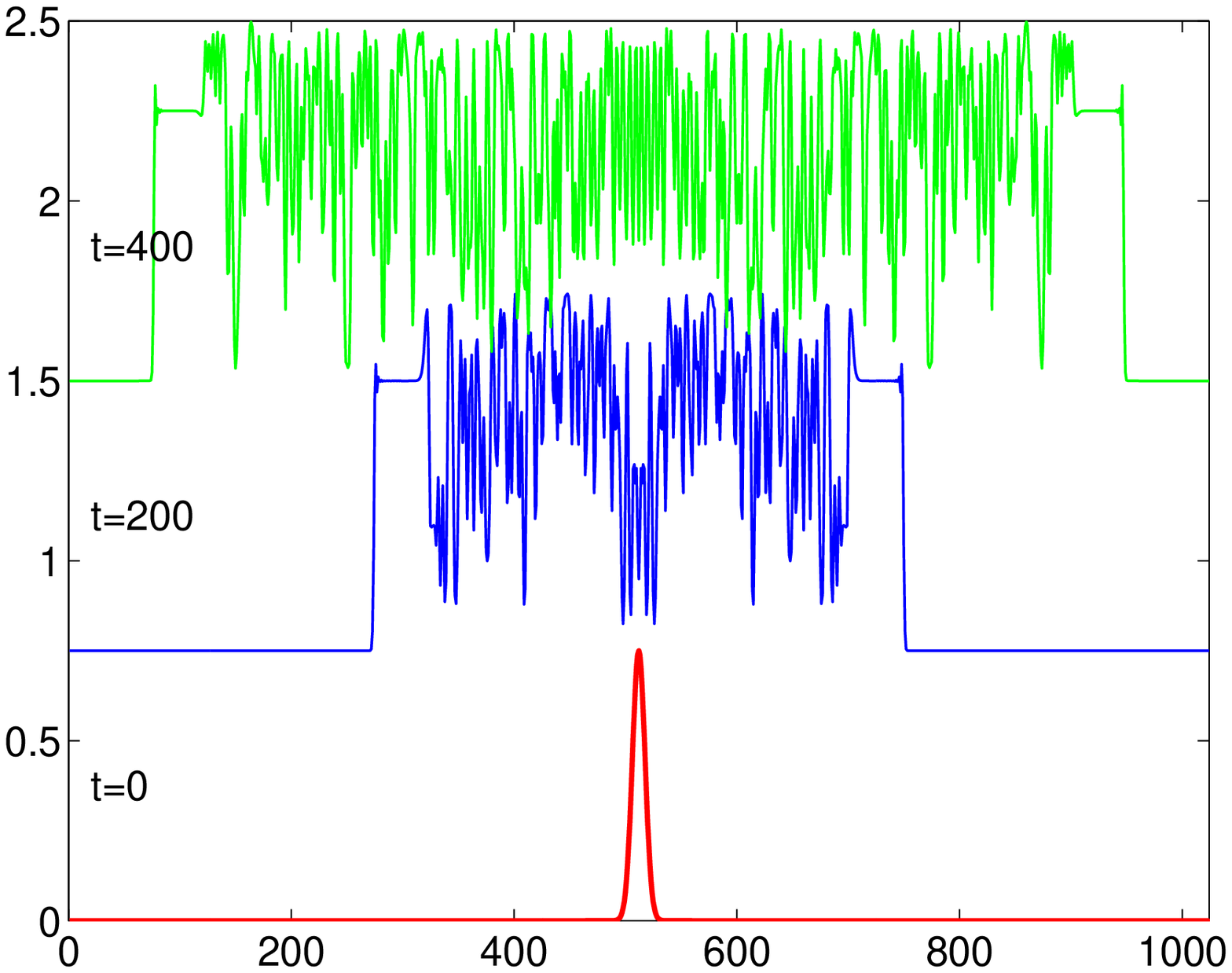}
\caption{Evolution of the initial configurations for (a) a step-like initial
condition and (b) a Gaussian bump.}
\label{travelling}
\end{figure}

In Fig.~\ref{travelling} (a) and (b) we show the evolution of the
CML starting from localized configurations with a
step-like initial condition in panel (a) and a Gaussian form in panel
(b). The configuration $x_n^i$ at three times, $n=0$, $n=200$ and $n=400$
are shown with different lines. They have been translated
vertically to render the identification of the curves simpler.  For
both initial conditions, after a very short transient the
configuration acquires rather sharp borders with a `random'-like form
within the step. More precisely, the value at the plateau close to the
borders is well described by the stable solution $x^i_n \sim
(r-1)/r=3/4$ while the values of the random-looking peaks on this
plateau have an average $\langle \, x \, \rangle \sim 0.67$ as in
Sect.~\ref{noise}. For these times the decay to zero at the edges is
very sharp, it occurs in less than 10 lattice spacings.

\begin{figure}[htb]
\includegraphics[width=7cm]{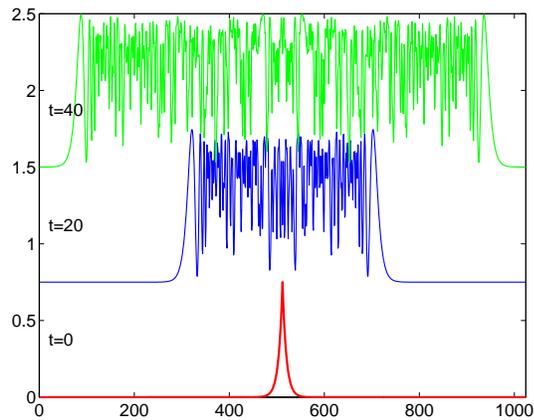}
\caption{Evolution of the initial configurations for an exponentially
decaying initial condition.}
\label{travelling-fast}
\end{figure}

In Fig.~\ref{travelling-fast} we display the evolution of the
CML starting now from a more extended initial
conditions; more precisely, we took an exponentially decaying form
$e^{-x/\xi}$ with $\xi=8$, and we show the configuration $x_n^i$
at three times $n=0$, $n=20$, and $n=40$. It is clear that the
front travels much faster in this case than in the examples
studied in Fig.~\ref{travelling}. Moreover, the interface is not
as sharp.

We delay for future work~\cite{Francesco} a quantitative analysis of
the velocity of propagation.  One might expect that additional
non-linearities, as the KPZ-like term, can only increase (or leave
unchanged) the velocity with respect to $v_{min}$.  A sufficient
condition to have a pulled front is that all non-linear terms should
suppress growth for a front that propagates into a linearly unstable
state. In our modified FKPP equation it is not obvious {\it a priori}
whether this holds.

\section{Conclusions}
\label{sec:conclusions}

We have shown that in the continuum space and time limits the CML of
logistic units in $d=1$ becomes an equation of the KPZ type.  We
argued that some non-linear terms have a triple effect. These terms
can be interpreted as a noise source with short-range correlations in
time and space, an assumption that we supported with the statistical
analysis of numerically generated configurations.  But they also
provide a source for confinement thus reducing considerably the
roughness of the surface: the positive roughness exponent $\alpha$
found in conventional KPZ is reduced to a vanishing value.  On the
other hand, the same non-linear terms have the form encountered in the
FKPP equation for the diffusion of advantageous genes and one can then
expect to find travelling wave configurations in the discrete
model. We also argued that the field-dependent viscosity appearing in
the partial differential equation has no special effect.

As regards to time-dependent observables in the CML,
we found that the correlations in time of the local fluctuations
have a stretched exponential decay as recently claimed to
arise for surfaces generated by the conventional (unbounded) KPZ
equation~\cite{Moore}-\cite{Eytan}.

We have shown results for the pair of parameters $\nu=0.4$ and $r=4$
where the system is deep in the chaotic regime. A detailed analysis of
the phase diagram is beyond the scope of this article. We just mention
that the stretched exponential decay persists when
keeping the value of $\nu$ fixed and reducing $r$ until reaching the
`critical' value $r\approx 3.83$. Interestingly enough, long periods
of local blocking are not necessary to find such a slow relaxation since
we find it even when the space-time plots look extremely chaotic. Time
correlations are hidden in these plots.

A number of authors have signalled the possible relevance of CMLs in
describing different aspects of glassy relaxation.  Recently, Mousseau
{\it et al}~\cite{Mousseau,Sim} studied the same model
in its intermittent regime ($\nu=0.4, \; r\approx
3.83$) with the aim of relating the 10 orders of magnitude stretched
exponential decay of its distribution of trapping times (times in
which an element remains locked into one of the coarse-grained values
$s_n^i=\pm 1$) to the one observed in super-cooled liquids. Simdyankin
and Mousseau ~\cite{Sim} associated this stretched exponential decay
to the one of the correlation function.  As we have already stressed,
we also obtain a stretched exponential relaxation for larger value of
$r$ where trapping intervals for the coarse-grained variables $s_n^i$
do not exist (see Fig.~\ref{figa}).  In a similar spirit to the works
of Mousseau, Garrahan and Chandler associated~\cite{Chandler} the slow
dynamics and glass transition to the structure of trajectories, in the
form of the space-time maps, of the spin-like variables in kinetically
constrained lattice gases (see~\cite{Peter} for a review of these
models). A fully-connected model of logistic maps was studied from the
glassy point of view by a number of authors; a range of values of $r$
with a large number of `macroscopic states' and `ergodicity
breaking'~\cite{vulpiani} and the possibility of an analog of
replica-symmetry-breaking~\cite{manrubia} were reported. With the
baggage gained from the current understanding of the dynamics of
glassy systems~\cite{Leshouches} we intend to revisit the generation
of an effective temperature~\cite{Shraiman-Hohenberg,Cross,Teff} and
its possible appearance in a fluctuation
relation~\cite{Francesco,Sasa} in chaotic systems~\cite{prep}.

Finally, we also briefly analysed the resulting partial differential
equation in comparison with the FKPP and we showed that the CML
has travelling wave solutions with similar qualitative properties to
the ones in FKPP. A more detailed analysis is necessary to understand the
main characteristics of the front and, in particular, classify
its velocity depending on the initial conditions and other parameters.

\vspace{1cm}

\noindent\underline{Acknowledgements} We thank H. Chat\'e, B.
Derrida, J. A. Gonz\'alez, J. Kurchan, S. Majumdar, N. Mousseau,
I. Procaccia, L. Trujillo  and F. Zamponi for very helpful
discussions.  L.F.C. is a member of the Institut Universitaire de
France.

\end{document}